# MarkupLens: Balancing Computer Vision Assistance and Control in Professional Video Annotation for Video-Based Design Tasks

TIANHAO HE, Delft University of Technology, The Netherlands

EVANGELOS NIFORATOS, Delft University of Technology, Netherlands

GERD KORTUEM, Delft University of Technology, The Netherlands

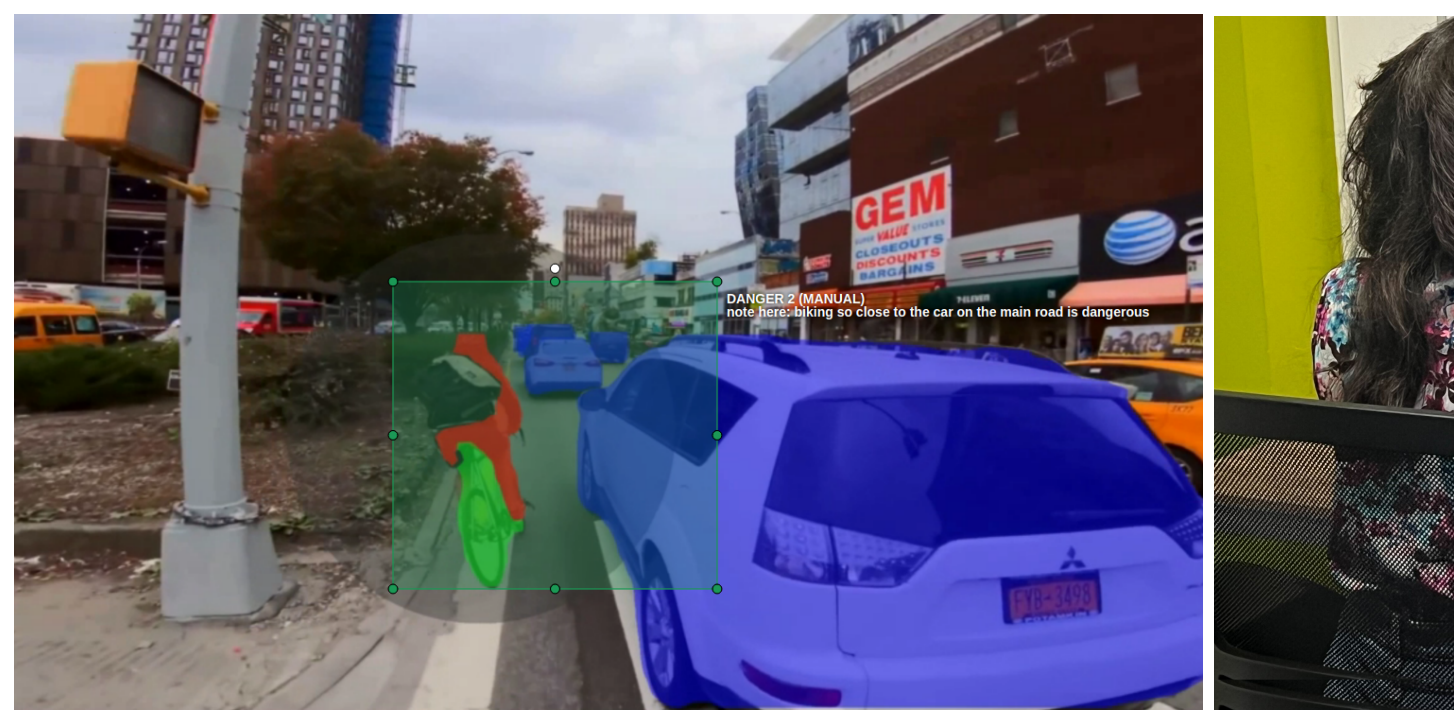

(a) A designer annotation using *MarkupLens* with automatic annotation support

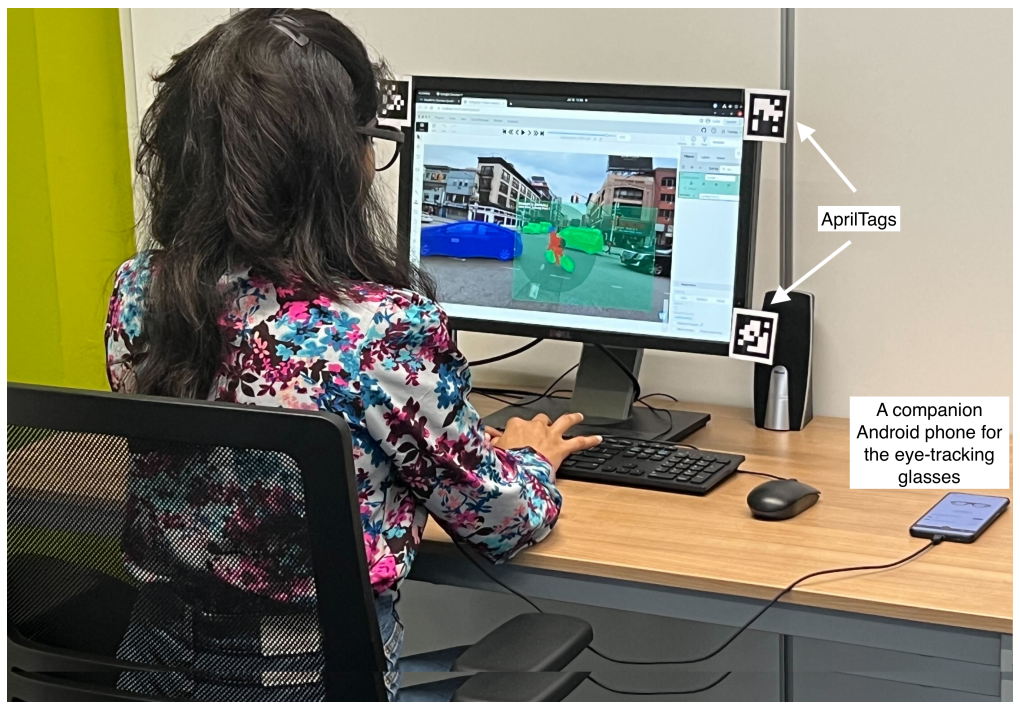

(b) The study environment setting

Fig. 1. *MarkupLens* is an Computer Vision (CV)-powered prototype that combines off-the-shelf detection models to track and automatically label relevant video in video-based design research. We evaluated *MarkupLens* with 36 designers in a VBD user study and in three automatic support: (1) "no", (2) "partial", and (3) "full" support.

Video-Based Design (VBD) uses video as a primary medium for analyzing user interactions, prototyping, and generating design insights. However, current VBD workflows are constrained by labor-intensive, inconsistent manual annotations that fragment attention and delay insights. Computer Vision (CV)-powered automatic annotation offers opportunities to reduce manual effort while supporting higher-level interpretation. This paper investigates human-AI collaboration in video analysis by examining how different levels of automated support shape user experience in VBD. We developed *MarkupLens*, a CV-assisted annotation platform, and conducted a between-subjects eye-tracking study with 36 designers in an urban VBD case. We compared three levels of automation: no support, partial support, and full support, and found that higher levels improved annotation quality, reduced cognitive load, and interestingly, enriched reflection. Our insights on automation levels inform adjustable autonomy and mixed-initiative system design beyond VBD tasks.

---

Authors' Contact Information: Tianhao He, Delft University of Technology, Delft, The Netherlands; Evangelos Niforatos, Delft University of Technology, Delft, Netherlands, e.niforatos@tudelft.nl; Gerd Kortuem, Delft University of Technology, Delft, The Netherlands.

## 1 Introduction

Video as a multi-sensory medium, typically carries rich information. Video can be employed to foster an enriched learning experience, engaging learners through a blend of visual, auditory, kinesthetic, and emotional components. In design, the adoption of video as a primary tool for knowledge generation, idea collation, and problem identification is known as Video-Based Design (VBD) [24, 62, 66, 69, 70]. The practice of VBD provides learning experiences and enables designers to gain a deeper understanding and relevant knowledge. For example, Mackay et al. integrated videos of user interactions into the re-design process of a graphical editing prototype [45]. They revisited these videos in details and brainstormed their ideas derived from the footage on a whiteboard to share with their colleagues. Traditional practices of note-taking for videos, such as transcribing observations and ideas onto whiteboards separate from the video medium, have been a go-to approach for many designers [69]. However, the process of dissecting videos can be inefficient and may occasionally overlook nuances [70].

The difficulties from the traditional methods become prominent when VBD practitioners need to make precise notes on zoomed-in scenes and synchronize them with specific frames of interest. The existing tools such as ELAN [1] and Transana[2] try to address these problems to provide note-taking software but fall short in addressing the iterative and exploratory nature of the VBD workflows. Similarly, existing AI-driven platforms focus on objective metrics such as pose estimation [13, 50] and overlook the subjective and context-rich insights essential to VBD. These limitations suggest the need for a more adaptable solution that enables designers to extract higher-level reflections from video in VBD practices.

Video annotation, defined as the process of adding labels and tags to recordings, has an alternative possibility to fill this gap. Video annotation includes marking elements with text, bounding boxes, and categories on the interested frames of videos. It is commonly performed at scale through crowdsourcing, which makes the workflow accessible for first-time users [10]. It also enables time- and location-precise note-taking with synchronization, which is essential for VBD practice [70]. Beyond manual annotation, the variations from Computer Vision (CV), such as object detection and semantic segmentation have become mature and widely accessible [73]. These algorithms can automatically detect objects of interest to offer practitioners relevant cues and reduce the manual effort involved in information retrieval.

Prior work primarily explored how CV augmentation impacts video annotations for training models [7, 25, 37, 73]. However, annotations in VBD contain more than just plain descriptive names of objects [39]. Moreover, given the distinct nature of design compared to traditional use of annotation, excessive automation in video analysis risks overlooking human insight and reducing opportunities for reflection [1]. Chen et al. advocated for improvements aimed at increasing the resilience of human decision-makers by enabling them to override automatic predictions when necessary [15]. Similarly, Buçinca et al. provided solutions aimed at reducing over-reliance on automation by adopting cognitive forcing interventions [11]. It is therefore crucial to understand how automatic annotation assistance can facilitate video analysis in design while preserving practitioners' autonomy. In this paper, we introduce *MarkupLens*, a prototype featuring three distinct levels of CV support in video annotation for VBD practice. Specifically, we examine: (1) no CV support, where designers manually annotate VBD videos; (2) partial CV support, where the system provides a limited subset of automatically generated annotations; and (3) full CV support, where all detected annotations are displayed automatically. These levels were chosen to represent a continuum of automation, allowing us to investigate how varying degrees of automatic assistance shape annotation practices in VBD tasks. Our work makes the following contributions:

[1] https://archive.mpi.nl/tla/elan (last accessed: November 19, 2025).
[2] https://www.transana.com/ (last accessed: November 19, 2025).

- An empirical study of levels of automation in collaborative video analysis. We compare three conditions: no support, partial support, and full support, and show how different levels of CV-supported automatic annotation assistance influence practitioners' VBD practices, workload, and experience.
- Evidence of how automation shapes VBD annotation strategies. Increased automation shifted practices away from producing many short notes toward fewer, longer entries, indicating a reallocation of effort and attention.
- Workload and user experience varied across levels of automation. Automation reduced perceived workload and improved user experience ratings. These gains, however, also raised design concerns about trust, reliance, and control.
- Implications for adjustable autonomy design in analysis tools. Beyond the design case presented here, our results inform the design of mixed-initiative systems where humans and AI collaborate to interpret complex data. They also speak to long-standing HCI discussions on how to scale support levels effectively.

## 2 Related Work

### 2.1 Video-Based Design

VBD is a methodology in design research that heavily relies on video as a tool for capturing, analyzing, and communicating information. This approach is particularly prevalent in fields such as UX and interaction design, and ethnographic research [70]. In VBD, videos are used to capture and study user interactions with products or environments, offering a dynamic and context-rich source of data. Designers analyze these videos to gain insights into user behaviors and needs, which in turn inform the design process [69, 70]. Additionally, videos can be created to demonstrate and test design concepts and ideas. The later use of VBD enables a deeper understanding of real-world user interactions and experiences. Such insights guide design practitioners toward more effective and problem-centered solutions. Tracing back to 1989, designers from Apple Computer Inc. (now Apple Inc.) leveraged videos to conceptualize new user interfaces (UIs) of upcoming computers [66]. Capitalizing on the temporal representations offered by videos, they created animated objects using computer programs within video productions. This approach enabled them to simulate new UIs, facilitating the internal dissemination and evaluation of their novel UI designs. In the same year, Tatar from Xerox PARC (now PARC) initiated an exploration into learning from repeated observations and documentation of user behavior through stationary camera recordings [62]. This approach was aimed at reducing erroneous assumptions in the development of new software. Tatar also emphasized the necessity for more advanced video editing tools for software designers as the video-based interaction analysis is a complex and multifaceted task [62]. Later, Ylirisku and Buur stated that VBD is useful for design practitioners to learn from users' daily experiences and acquire "thick descriptions" of an artifact. Videos are instrumental in capturing users' movements, interactions, and emotional transitions, thus constructing design narratives and encapsulating individual thoughts for further examination by designers [69]. The VBD approach aids in comprehensively understanding which aspects should and should not be improved in design prototyping over time [70]. Similarly, the concept of learning from videos, known as Video-Based Learning (VBL), has been adopted to enhance understanding for knowledge workers in various domains including primary care and education [2, 5]. Designing from videos often requires note-taking with precise timing in the video and indications on the location of the view, which may take a huge amount of time and effort for designers to locate, process, record, and reflect upon during the process. This challenge can be potentially addressed by incorporating annotation techniques from crowdsourcing.

### 2.2 Video Annotation Techniques and Tools

Video annotation is a process of adding text labels, bounding boxes, and other forms of tags to video frames to categorize objects, events, or features. To improve efficiency and accuracy, tools combining automatic assistance with human annotators have become prevalent in recent years for labeling objects in video footage [25]. Bianco et al. developed an interactive video annotation tool that combines manual, semi-automatic, and fully automatic annotation using CV algorithms such as linear interpolation, template matching, and supervised object detection with Haar-like, HOG and LBP features [7]. Their tool also incorporates incremental learning and object tracking to reduce human effort and enhance scalability. Bianco et al. demonstrated that cases utilizing semi-automatic and fully automatic CV support reduce human effort while maintaining the quality of the annotations. Similarly, Kavasidis et al. proposed a collaborative web-based video annotation tool with CV algorithms such as GrabCut, Snakes, and Canny aimed for contour extraction in collecting video ground truth annotations at scale with efficiency and quality in mind [37]. Another case in point is Mu, who introduced annotation software specifically designed to combine classroom videos with transcriptions for a learning framework. The author provided a comprehensive demonstration of how users can effectively leverage the timestamp-based hyperlinks in their system to meticulously document text descriptions, supplemented by time anchors, within the log history [49].

However, existing video annotation tools are predominantly designed for general-purpose tasks and lack features that cater to the needs of VBD. Tools like OpenPose [13] and DeepLabCut [50], for instance, excel at objective motion tracking and pose estimation but are limited to labeling simple objects or marking keypoints for training detection models. OpenPose focuses on detecting and tracking human skeletal structures with precision but does not provide support for subjective or contextual insights for reflective design processes [13]. Similarly, DeepLabCut is optimized for animal tracking and motion analysis which is employed algorithms for high-accuracy detection, yet it remains narrow in scope of offering no support for iterative workflows or subjective annotations [50].

Other tools, such as ELAN [3] and Transana™ [4], emphasize iterative annotation and qualitative analysis, often used in linguistic and behavioral research. While these tools offer robust capabilities for transcription and coding in videos, they lack CV integration to automate routine tasks or enhance scalability, which limits their possible application to extensive and iterative VBD workflows. Additionally, their rigid interfaces do not support collaborative or real-time annotation, further constraining their utility in dynamic design settings. Collaborative platforms, for example, Encord [5] and V7 [6] have introduced team-based annotation features and multi-view analysis which facilitate primarily to the area of AI dataset creation. These tools integrate automation for object tracking and bounding box generation but fall short in supporting subjective interpretation and exploratory workflows in VBD. Their focus on CV pipelines often sidelines the creative and reflective processes designers require for exploring in videos [70]. In this paper, we introduce *MarkupLens*, an CV-powered prototype specifically tailored for the complex and reflective processes of VBD practices. We evaluate *MarkupLens* in a VBD user study to address gaps in iterative workflow support. Our prototype emphasizes subjective exploration and reflections in VBD, which are features that existing tools fail to deliver.

The CHI community has shown interest in exploring the potential use of video annotation to address diverse challenges across domains. For instance, Zhang et al. proposed *RCEA*, a user-centered annotation tool that uses emotion annotations to capture behavioral interactions in mobile videos. It reduces challenges associated with multitasking and mental workload [72]. Similarly, *EventAnchor* [20] features scene detection and object tracking to improve the efficiency of annotating sports videos with less manual effort. Building on automation, *TmoTA* [52] introduces a responsive manual labeling tool to speed up object tracking annotations and ease user workload. In the medical field, *Surch* [38] generates annotated procedural graphs for surgical videos and supports video navigation of treatment phases. In the context of supporting blind people, *Ga11y* [71] combines machine learning-based visual similarity matching with human annotations to enhance accessibility and usability of short videos. The versatility of video annotation tools highlights their adaptability across diverse contexts. However, their potential to augment tasks in design has received little attention, which presents a valuable opportunity for further exploration.

[3] https://archive.mpi.nl/tla/elan (last accessed: November 19, 2025).
[4] https://www.transana.com/ (last accessed: November 19, 2025).
[5] https://encord.com/ (last accessed: November 19, 2025).
[6] https://www.v7labs.com/ (last accessed: November 19, 2025).

### 2.3 Computer Vision in Video Annotation

CV is a subset of AI that enables computer programs to extract features and derive information from visual content. Its objective is to mimic the visual perception capabilities of humans, enabling it to make recommendations based on visual cues. Supported by Deep Learning (DL) algorithms, CV has experienced exponential growth in recent years. It now finds application in video annotation in a wide array of areas, tackling tasks such as classification, detection, and segmentation across diverse visual contexts, including static images, moving videos, and zoomed-in scenes [67]. In recent years, professional annotation tools employing CV have also experienced a transformation. Such tools have evolved from serving general purposes to addressing domain-specific challenges. In the field of agriculture, Rasmussen et al. presented a specialized Faster Region-Based Convolutional Neural Network (Faster R-CNN) integrated with a semi-automatic annotation platform to help researchers identifying corn fragments [53]. Furthermore, the MarkIt annotation platform, as proposed by Witowski et al., simplifies the annotation process for radiologists when interpreting and analyzing x-ray images. They employed supervised learning to categorize symptoms by classes. Their tool reduces the effort required from human experts in medical annotation processes [68]. In manufacturing, Liang et al. developed a causal CV method that combines domain-specific reasoning to address challenges in vision-based quality inspection [40]. Their approach improves defect detection accuracy and reliability by mitigating visual interferences such as reflections and overlapping components in complex industrial scenarios. Similarly, Ding et al. developed a cognitive visual inspection system for screen manufacturing with deep-convolution neural networks to enhance defect detection and classification accuracy by addressing uneven lighting, reflections, and defective surfaces [21]. Beyond manufacturing, CV and annotations have significantly impacted the energy industry. For example, Nguyen et al. proposed an autonomous vision-based power line inspection system that utilized drones and CV algorithms to identify defects in power line components and improve inspection speed and accuracy [35]. While CV has proven effective in domains ranging from agriculture and medicine to quality control in manufacturing and the energy industry, its potential to augment the reflective and context-sensitive tasks of design with videos remains unexplored.

Building on recent advancements in CV and annotation techniques, there is an opportunity to address existing challenges in current VBD practices. **We introduce an CV-powered augmented video annotation prototype (*MarkupLens*) to enhance VBD tasks. In our study, we investigate the impact of *MarkupLens* on participants' performance and mental workload in a scenario-based video design case across three distinct levels of automatic assistance in VBD video annotation: (1) "no", (2) "partial", and (3) "full CV support".**

## 3 The *MarkupLens* Prototype

In VBD, designers rely on video as a medium to capture and reflect on their design findings [70]. This requires tools that can effectively support the analysis of design insights derived from visual cues, such as user interactions or spatial

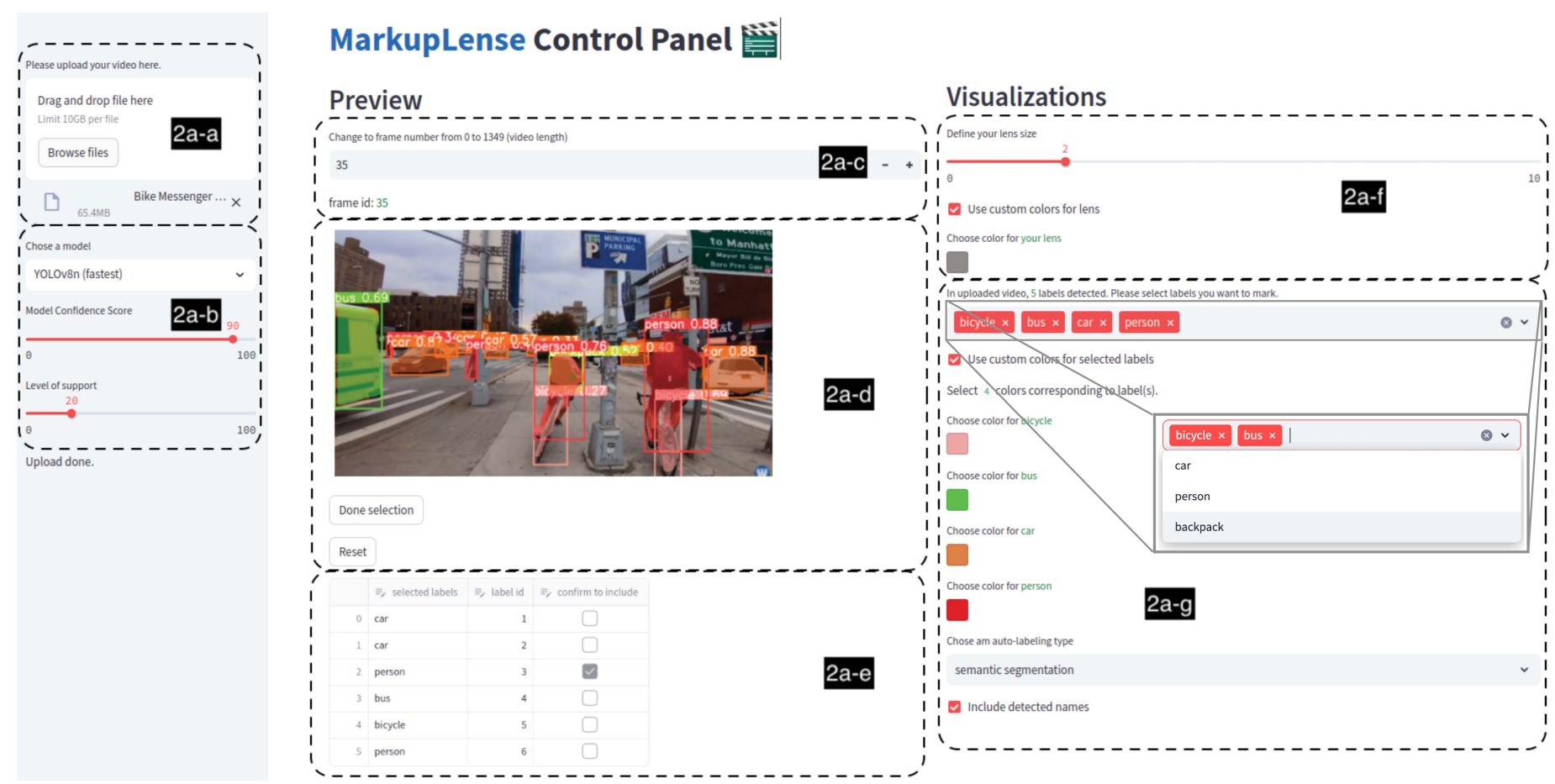

(a) A screenshot of the Control Panel showing a biker on the pavement: Region 2a-a enables video uploads; Region 2a-b provides options for selecting video processing models, setting object detection confidence, and adjusting automatic support levels; Region 2a-c offers a frame selection for video preview; Region 2a-d displays the preview frame, allowing users to choose instance(s) for tracking within the "lens" as a focus (foci); Region 2a-e contains a confirmation list to review selected instances for the "lens" in case multiple instances are selected; Region 2a-f facilitates setting the radius and color of the "lens", where selected objects from Region 2a-g are displayed in the "lens" area with specified colors and a visualization method. A zoomed-in window is also shown to demonstrate how the potential labels are listed for designers in Region 2a-g.

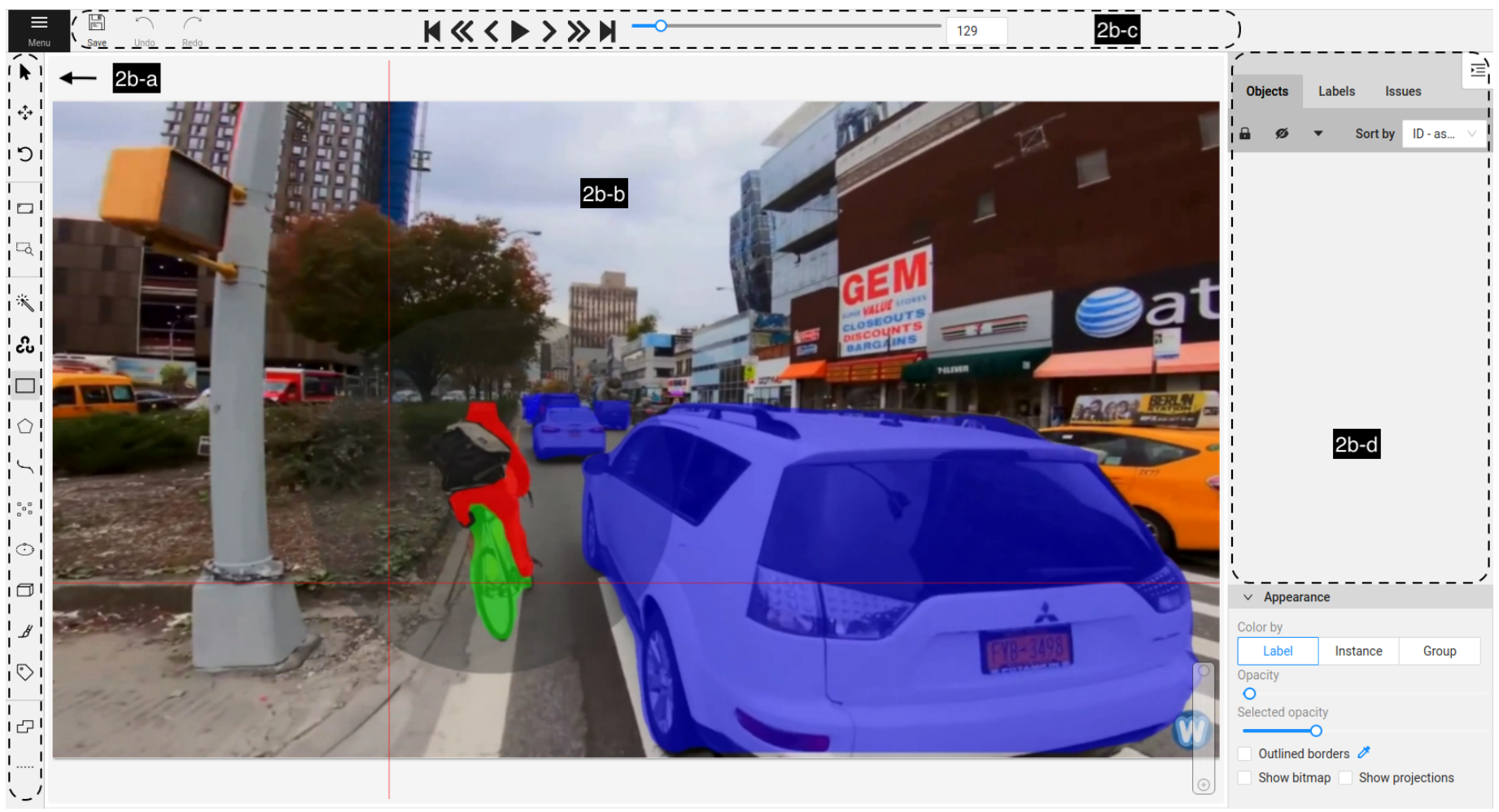

(b) A screenshot of the Annotation Page showing a biker riding in traffic: Region 2b-a contains a sidebar with options for creating annotations; Region 2b-b shows as the video viewing window; Region 2b-c contains video player controls, enabling users to play, pause, and navigate video progress; Region 2b-d is the note window, where users can view and edit existing annotations.

Fig. 2. The *MarkupLens* Prototype: Control Panel and Annotation Page in a video-based urban design case

dynamics observed in videos [69]. To meet this need, designers require annotation systems that combine accuracy with the ability to filter information and retain relevant elements. Existing annotation tools offer precise object detection and labeling but lack the flexibility for customized VBD annotations.

To address this gap, *MarkupLens* introduces an innovative approach that combines precision and customization tailored specifically for designers' needs in VBD. The *MarkupLens* includes two interconnected web-based UIs: a control panel (Fig. 2a) and an annotation page (Fig. 2b). Starting with the control panel, we integrated 7 variations of foundation models from YOLOv8[7], a state-of-the-art (SOTA) model frequently utilized for object detection and instance tracking. Users can upload their videos (2a-a), choose an object detection model with varying levels of automatic support (2a-b), and customize the range and type of visualizations (2a-c to 2a-g). Specifically, in region 2a-b, the control panel includes two sliders: model confidence score and level of automation support on annotation. The model confidence score lets designers adjust the sensitivity of video-based automatic detection, ranging from 0 (no confidence) to 100 (full confidence). A higher confidence score results in a more sensitive model, capable of detecting a greater number of objects categorized by the designer, but with a potential trade-off of miscategorizing objects due to lower accuracy threshold. Conversely, a lower score increases detection precision at the expense of potentially missing some relevant objects. As such, designers can leverage the tunable capabilities of SOTA foundation models and adjust it based on their personal preferences and specific needs in the design practice. The score further provides designers with an intuitive quantitative estimate of the validity from models' predictions. The second slider ("level of support") controls the extent of model assistance in VBD tasks. This option is designed to mitigate the potential for over-reliance on automation, which can limit reflective design exploration [41]. It allows designers to adjust the level of computational assistance and promotes a balance between automated support and human input for a more flexible design process. In addition, the region 2a-c in Fig. 2a prompts users to select a frame containing the instances they wish to analyze in their tasks. The chosen frame is displayed in region 2a-d, where users can preview the current selections and choose on the interested object(s), such as a biker as exampled in Fig. 2a in an urban design task. If multiple objects overlap at the click point(s), the interface generates a list in region 2a-e for double confirmation. Users can then check boxes in the "confirm to include" to select and track one or multiple instances depending on their needs.

In region 2a-f, users can define the automatic detection range (referred to as the "lens") and assign it a favorable color. This "lens" addresses limitations of general-purposed annotation tools by allowing designers to intentionally focus on areas of interest and filter out irrelevant details from VBD tasks. Subsequently, in region 2a-g, a list of detected instances appears (see zoomed-in window in Fig. 2a), allowing users to select their interested elements from a drop-down list. Designers can also select different modes of automatic annotation, such as boxed object detection, image recognition, instance and semantic segmentation (see Fig. 8 in appendix for examples). Designers can iteratively repeat this process of selecting frames and assigning labels if they find elements (e.g., road signs in an urban deign case) are not included in the label list in region 2a-g. Once the selections are finalized, the control panel processes the video depending on the settings collected and transfers to the annotation page. Since design is inherently personal, with preferences and working styles varying widely among individuals [69], the process is crucial because it combines the accuracy of detection models with the flexibility for designers to tailor selections to their specific needs from VBD tasks. Fig. 3 illustrates the overall workflow and showcases how designers interact with the control panel to configure and apply automatic annotations on VBD videos.

[7] https://ultralytics.com/yolov8 (last accessed: November 19, 2025).

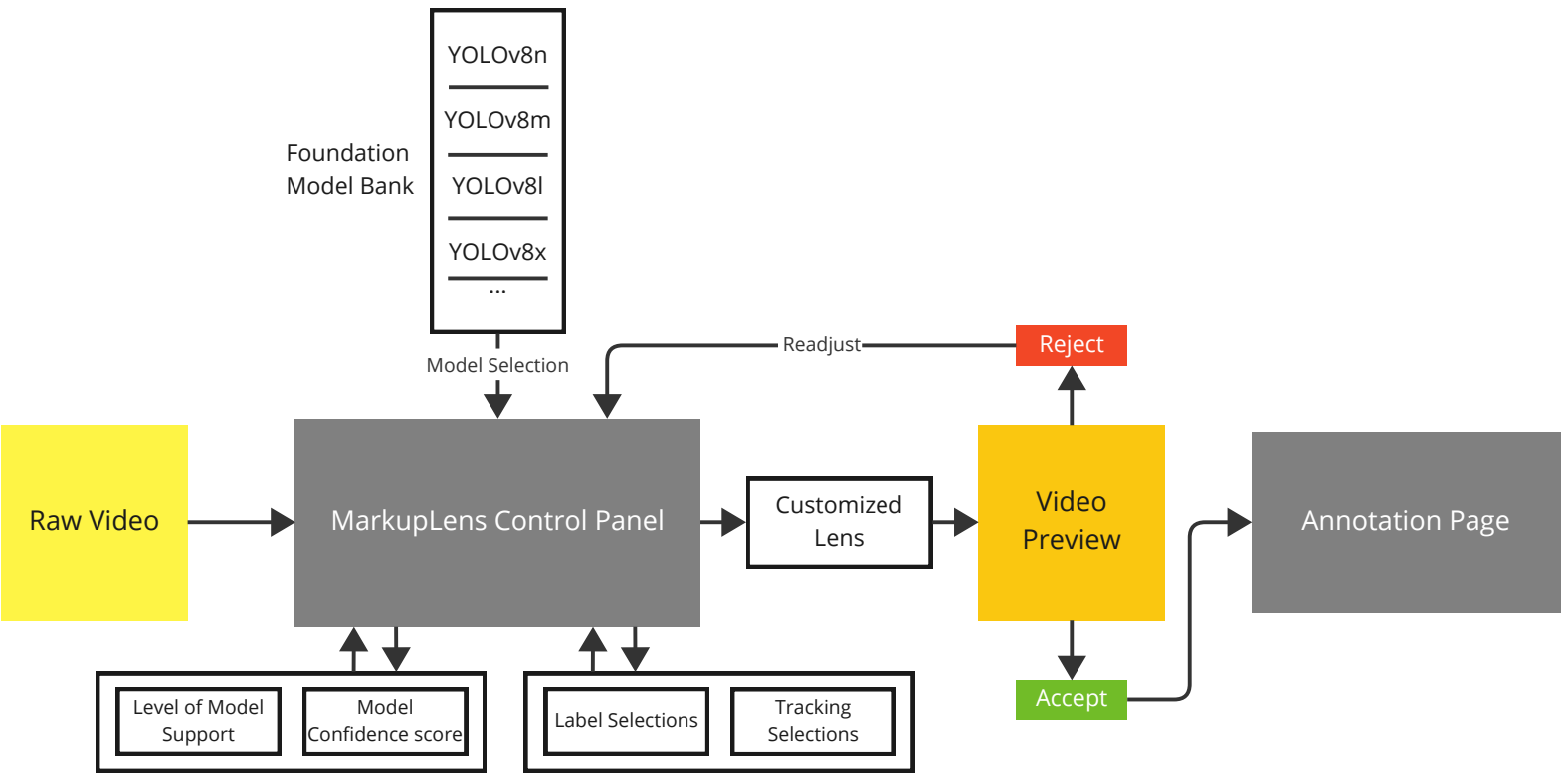

Fig. 3. A system workflow of *MarkupLens*

In addition, the annotation page (Fig. 2b) is a refined version inspired by a commonly used annotation tool CVAT[8]. In the side toolbar (region 2b-a), we simplified the original layout by retaining only one option (rectangle bounding box) for manual annotation in VBD tasks. This option provides a straightforward way to tag zoomed-in scenes. The video player bar, located at the top in region 2b-c, allows users to pause, play, reverse, accelerate, or navigate to a specific frame in the video. The central region 2b-b displays a biking scene from a video with automatic annotation support in an urban design case. Notes synchronized with frames and locations are shown on the right in region 2b-d. Users can then browse them frame by frame or with keyboard arrows.

## 4 Study

We assess the effectiveness of *MarkupLens* and the impact of automatic assistance on practitioners' design workflow by examining three levels of automation (**"no"**, **"partial"** and **"full"**) in an urban design case. We seek to answer the following Research Questions (RQs):

RQ1. *How do different levels of automatic assistance impact the quality of annotations in VBD tasks?*
Previous studies have highlighted many advantages of visual automation, particularly instance detection, in supporting professional cognitive tasks across diverse industries [53, 68]. However, none of these studies aimed for design-centric annotations on videos, nor explored how certain leveled automation can influence the quality of design thinking through these annotations. To address this gap, we implemented three levels of CV-supported automatic annotation in this study and introduced a design use case. Our hypothesis is that higher levels of automatic support will result in an **increased in the quality of design annotations** in the VBD process. In turn, these enhanced annotations can generate new insights beyond traditional evaluation methods and scale up the VBD process with CV.

RQ2. *How do different levels of automatic assistance influence the cognitive load that designers exhibit in the design process?*
Cognitive load refers to the demands placed on an individual's working memory by a given task [3, 4]. As indicated by prior research, a negative correlation is found between cognitive load and decision-making performance [22, 32]. We hypothesize that by incorporating a higher level of automation assistance in *MarkupLens*, we can reduce the cognitive load on design practitioners, which in turn could enhance their decision-making capabilities. Utilizing a combination of subjective methods, such as self-reporting questionnaires (**NASA-TLX** [34]), and objective measures, such as **eye-tracking**, we aim to provide a nuanced assessment of cognitive load across the three conditions.

[8]https://github.com/opencv/cvat (last accessed: November 19, 2025).

RQ3. *How do varying automation levels in MarkupLens affect designers' user experience (UX) and technology acceptance in VBD?*
Evaluating UX is crucial as it provides valuable insights into designers' views on a system. This evaluation is fundamental in the assessment of our prototype. We hypothesize that the integration of automatic annotation using CV will ease the video analysis process in VBD tasks. To glean deeper insights into how designers perceive the prototype, we employ a User Experience Questionnaire (**UEQ**) [58, 59] and a Technology Acceptance Model (**TAM**) (see section 4.3). We also hypothesize that higher levels of automation support in *MarkupLens* will enhance UX and that our prototype will be positively perceived in terms of technology acceptance, eventually being perceived as beneficial in VBD.

### 4.1 Participants

We recruited 36 design graduates (19 female and 17 male) from the Design Faculty of our University. Our participants are university students (BSc & MSc) and PhD candidates with an average age of 26.75 years (*SD* = 5.06) and average design experience of 2.86 years (*SD* = 1.13). Table 1 shows the participants' current educational level, their self-rated experience with design-oriented video annotation tools, and proficiency in using CV-powered tools (e.g., Midjourney[9]). Participants with a visual acuity below 20/20 were instructed to wear contact lenses prior to the experiment. All recruitment and study procedures were approved by the Research Ethics Board of our university, and informed consent was obtained from all participants. Before assignment, participants completed a demographics questionnaire including age, gender, education, prior video-annotation experience and CV-tool proficiency. We used stratified block randomization to distribute participants evenly across the three conditions (n = 12 per group). Post-assignment balance checks indicated no significant between-group differences on age (ANOVA $F_{2,33} = 0.711$, $p = .499$), gender (Chi-square $\chi^2(2) = 0.253$, $p = .895$), education level (Chi-square $\chi^2(4) = 0.268$, $p = .792$), video-annotation experience (Kruskal-Wallis $\chi^2(2) = 1.024$, $p = .539$) or CV-tool proficiency (Kruskal-Wallis $\chi^2(2) = 0.898$, $p = .638$).

### 4.2 Apparatus

We assessed *MarkupLens* in a typical office environment where the system was run locally on a desktop machine. Figure 1b shows the lab setup where participants interacted with the system. In addition to standard office tools such as a keyboard, mouse, and speaker, participants were instructed to wear eye-tracking glasses (NEON by Pupil Labs[10] with a sampling rate of 200 Hz), which were linked to a companion Android phone via a USB type-c cable for recording eye-movement data. The collected eye-movement data includes eye gaze trajectories, gaze fixation positions, and blink patterns. We positioned the monitor 55 centimeters away from the user-facing edge of the table. The 22-inch monitor was tilted and its height adjusted to be approximately 15 degrees below the participants' horizon line. This arrangement ensured a comfortable visual zone for the participants, spanning from -15 degrees to 15 degrees [28]. Additionally, we affixed four AprilTags[11] to each corner of the screen (see Figure 1b) to enable the eye-tracking glasses to identify the screen's surface.

[9]https://www.midjourney.com/home (last accessed: November 19, 2025).
[10]https://pupil-labs.com/products/neon/ (last accessed: November 19, 2025).

Table 1. Participants' design-related demographic details: values are presented as answer frequencies (f), followed by the corresponding percentages (%).

| Variable | Answer | f | % |
|---|---|---|---|
| Current educational level | Bachelor | 11 | 30.56% |
| | Master | 22 | 61.11% |
| | PhD (ongoing) | 3 | 8.33% |
| Design-oriented video annotation experience | Definitely not | 2 | 5.56% |
| | Probably not | 6 | 16.67% |
| | Might or might not | 3 | 8.33% |
| | Probably yes | 17 | 47.22% |
| | Definitely yes | 8 | 22.22% |
| Proficiency in using CV-powered tools | Never used before | 5 | 13.89% |
| | Beginner | 20 | 55.56% |
| | Intermediate | 11 | 30.56% |
| | Expert | 0 | 0% |

## 4.3 Measures

### *4.3.1 Objective Measures.*

- **Number of Annotations**: the number of annotations made on frames of the video during the VBD task.
- **Average Word Count**: the averaged number of textual words per description.
- **Completion Time**: task completion time for the VBD task in seconds.
- **Eye-Gaze**:
  - **Fixation Rate and Duration**: the rate and duration during which the eyes stay still in one position [33].
  - **Saccade Rate and Duration**: the rate and duration of rapid, conjugate movements of the eyes from one fixation point to another [54].
  - **Blink Rate and Duration**: the rate and duration of rapid closing and opening of the eyelids [46, 61].

### *4.3.2 Subjective Measures.*

- **NASA Task Load Index**: a questionnaire evaluates cognitive load across six attributes: mental, physical, temporal demands, performance, effort, and frustration [3, 4, 22, 29, 32, 34, 36, 43].
- **User Experience Questionnaire**[12]: a questionnaire used for measuring User Experience (UX) in interactive products. It has a benchmark method that categorizes raw UEQ scores into dimensions such as efficiency, perspicuity, dependability, originality, and stimulation [58, 59].
- **Technology Acceptance Model**: a model for understanding user acceptance of new technological products and prototypes [17, 18, 65]. Based on literature, we modified statements from the original TAM (see Table 3 in appendix).

[11] https://april.eecs.umich.edu/software/apriltag (last accessed: November 19, 2025).
[12] https://www.ueq-online.org/ (last accessed: November 19, 2025).

### 4.4 Procedure & Task

*4.4.1 Preparation.* For experimental control, we selected foundational model YOLOv8x as our prediction model. Participants were given an urban design task based on a video[13] of a biker's commute through city streets. The edited video contained 5,400 frames (see video screenshots in Fig. 9). The completion of this task required approximately 10 minutes. We selected this duration to manage session timing and to align with prior VBD studies that commonly uses short video segments (5–15 minutes) for design analysis and prototyping [31, 44, 51]. **Our aim was to provide a standardized VBD task for participants to design safer biking commuting policies.** We configured the system to assign a fixed color to each automatically detected traffic entity, such as bikes, cars, and pedestrians. We also ensured that the biker in the video remained at the center of the "lens," defined as the automatic detection range (Fig. 2b; see Sec. 3 for details). The lens was set to a predetermined size and colored grey. We used instance segmentation to visualize annotations with an alpha value of 0.8 for color saturation.

Thirty-six participants were assigned to three groups in a between-subjects design. Each group consisted of 12 participants. Group (1) received **full CV annotation** with all detected annotations. Group (2) received **partial CV annotation** with 50% of randomly selected annotations per category. Group (3) had **no CV support** and analyzed the video independently. All participants were individually invited to the lab according to their scheduled times. They were first briefed on the purpose of the experiment and then seated in front of the monitor (see Fig. 1b). Before proceeding, participants were screened to ensure sufficient visual acuity for wearing the eye-tracking glasses.

*4.4.2 Training and Main Session.* After preparation, participants watched a three-minute training video demonstrating how to use the prototype. The training video instructed participants on using the playback features of the integrated video player and on basic annotation skills, such as adding text descriptions and drawing boxes on video frames. Subsequently, they were given sufficient time in a hands-on practice task to ensure they understood how to operate the UI. The training procedure took about 15 minutes. Next, participants were guided to wear eye-tracking glasses described in Section 4.2 and instructed to start the main task. Timestamps were recorded at both the start and end of the tasks. Within the annotation page, the video window (region 2b-b in Fig. 2b) was designated as the Area of Interest (AOI) for eye movement detection. **We introduced our participants to the predefined design case, which involved creating safer commuting policies for bikers. They were instructed to view and annotate a cycling video on *MarkupLens* as VBD practitioners. Their task was to generate annotations and text descriptions on events, interactions, or scenes for analyzing the given VBD video. Participants were advised to perform as naturally as they would in their regular VBD tasks and were given the flexibility to take as much time as needed for the task.** After completing the main task, participants were prompted to assess their workload using the NASA-TLX and provided feedback on their overall experience by completing both the TAM and UEQ surveys.

*4.4.3 Post-Session.* After completing all tasks, we inquired about their overall experience, as well as any other concerns or questions they might have had about the study. We then appreciated the contribution made by our participants and provided each of them with a 10-credits gift card.

[13] https://youtu.be/yWHdkK5j4yk?si=fuvizCEJoBeA3B3B (last accessed: November 19, 2025).

Table 2. Annotation measurements across three designer groups with different CV supports: values are presented as means (M), followed by the Standard Deviations (SD) in parentheses. Values with significant differences are marked with asterisks (*).

| Measures | No CV | Partial CV | Full CV |
|---|---|---|---|
| Nr. of Annotations (N) | 26.33(5.32) | 15.08(2.09) | 12.83(6.42)* |
| Average Word Count (N) | 2.23(1.65) | 5.02(2.044)* | 6.68(3.05)* |
| Time Spent (seconds) | 623.08(509.13) | 732.06(446.12) | 583.73(360.89) |

## 5 Statistical analysis and results

### 5.1 Quantitative Measurement of Annotations (RQ1)

We conducted a quantitative measurement of annotation in the study across three groups ("no", "partial", and "full CV support") as shown in Table 2. We observed a significant main effect between "full" and "no CV support" groups on the average number of annotations produced by designers (Kruskal-Wallis $\chi^2(2) = 6.24$, $p = .04$), with a mean rank number of 13.71 for "full CV support" group and 24.29 for "no CV support" group. Moreover, we noted a significant effect in the average word count per annotation (see second row in Table 2) across the three groups (ANOVA $F_{2,33} = 11.25$, $p < .001$, $\eta^2 = .41$). **Designers with "partial" and "full CV support" produced on average significantly more design insights in words (see Table 2) compared to those with "no CV support"** (Bonferroni $p = .02$; $p < .001$). Additionally, there was no statistically significant difference noted between the "partial" and "full support" in terms of average word count per annotation (Bonferroni $p = .27$). Likewise, no significant effect was identified in the time spent on tasks across the three designer groups (Kruskal-Wallis $\chi^2(2) = 1.21$, $p = .55$).

### 5.2 Qualitative Analysis of Insights from Annotation Text (RQ1)

Given the findings in Section 5.1 and Table 2, which showed that VBD practitioners made fewer but longer annotations as automatic support increased, we conducted an additional qualitative analysis of the annotation text. We then analyzed the purposes of the annotation texts made by participants and categorized them into four subsets (Sections 5.2.1, 5.2.2, 5.2.3, and 5.2.4) based on their length and content.

*5.2.1 Objective names.* Participants who received less automatic support from CV tended to leave a higher number of annotations with fewer descriptive words in each. Annotations in this category often included only the names of single objects that designers observed in the video context. For example, participants P2, P10, P16, and P18 from the "no CV support" group often left annotations such as "vehicles," "cars," "big trucks," "people," "taxi," and "van." Participants in this subset tended to make short, repetitive annotations across multiple related scenes. Although similar patterns were occasionally observed in participants from the "partial" group (n = 5; e.g., "unaware" and "children" from P26), this behavior was more frequent in the "no CV support" group (n = 12). Fewer participants in the "full CV support" group (n = 2) showed this pattern.

*5.2.2 Deeper contextual descriptions.* Designers who tended to write longer text included more descriptive annotations rather than only the names of objects. These descriptive annotations were often observed in groups with "no" (n = 5) or "partial CV support" (n = 3). Participants P5, P7, P11, P17, and P26 left descriptive annotations such as "walking people pass red traffic light," "cars coming from opposite direction," "car merging lanes," "red light is on," and "too close to the cyclist." These annotations contained more details about the task scenes than simple object labels. This level of detail may support reflection in a design process [69, 70]. Similarly, P13, P25, and P31 generated annotations such as "This car changes the direction," "Obstacle ahead," "black car on brakes," "a turning black car," "sudden close to people," and "squeeze into too narrow space." An interesting observation is that participants in the "no CV support" group (the group with no automated annotation support) often made more repeated annotations (n>3) in adjacent video frames or within the same scene compared to "partial" and "full CV support" groups (groups with 50% and 100% automatic annotation).

*5.2.3 Explanations on events.* We observed that designers in "partial" and "full CV support" groups focused more on explaining and hypothesizing possible reasons for the events observed in the given context. For example, P8, P11, P14, P19, and P32 annotated "no speed down at the zebra line and some people are too close", "the cyclist is in the blind zone for the driver of the van", "potential risks, cars may not decrease the speed because it might be green light", and "crashing into pedestrians because red stop sign was not respected". Participants in the "partial" (n = 7) and "full CV support" (n = 10) groups described the composition and configuration of the events more frequently in annotations compared to participants in "no CV support" group (n = 2). They also more often provided analytical insights by identifying potential consequences of decisions and explaining why the scenes might occur.

*5.2.4 Emotions and critical insights.* Many designers (n = 7) in the "full CV support" group demonstrated engaged emotions and critical thinking with the scene when automatic annotations of interest were provided. For example, P3, P12, P24, and P33 included text with exclamation and question marks in their annotations, such as "he scared the teenager!!!," "could hit a pedestrian with her dog! dangerous," and "riding against the cars???!!!". Based on the scenes shown to these participants, they also exhibited a high level of engagement by thinking critically and proposing suggestions. For instance, P4, P9, and P12 provided critical observations and potential suggestions, such as "The man is in the middle of the car path, which is too dangerous because he needs to pay attention to cars driving from two directions," "What if this black car is driving very fast? The man is interrupting its path," "He could have been hit by the car here. It's not safe to overtake between two cars," and "Overtaking the person from the left is dangerous." Most annotations (n = 35) in this category were from designers in the "full CV support" group, with several (n = 6) also from the "partial CV support" group. No such annotations in this subset (n = 0) were found in the "no CV support" group.

### 5.3 Cognitive Load (RQ2)

*5.3.1 NASA-TLX.* Prior literature shows a negative correlation between cognitive workload and users' decision-making capabilities in task performance [22, 32]. Accordingly, we assessed the cognitive load of the three VBD groups ("no," "partial," and "full CV support"). Figure 4 presents the TLX subscale scores and the weighted overall scores across the six attributes. The follow-up pairwise Mann-Whitney tests revealed that **designers in both the "partial CV support" group and "full CV support" group recorded significantly lower overall TLX scores compared to those in the "no CV support" group in the given VBD task** ($U = 7.50$, $p < .001$; $U = 5.00$, $p < .001$; "no CV support" group: $Mdn$ = 58.40, IQR = 14.94; "partial CV support" group: $Mdn$ = 42.02, IQR = 12.62; "full CV support" group: $Mdn$ = 32.74, IQR = 15.18). No significant differences were observed in the overall NASA–TLX scores between the "partial CV support" group and the "full CV support" group (Mann-Whitney $U$ = 50.00, $p$ = .22). In subscale TLX scores across 6 subsets, we observed a significant effect from *mental demand* in the three groups (Kruskal–Wallis $\chi^2(2) = 11.00$, $p = .004$). **Groups with "no CV support" reported higher mental demand scores compared to the "partial CV support" and the "full CV support" groups** ("no CV support" group: $Mdn = 57.21$, IQR = 18.93; "partial CV support" group: $Mdn = 33.20$, IQR = 21.24; "full CV support" group: $Mdn = 23.44$, IQR = 11.04; Mann-Whitney $U$ = 37.00, $p$ = .05; $U$ = 16.00, $p < .001$). For *physical demand*, we also observed a significant main effect in self-reported scores (Kruskal–Wallis $\chi^2(2) = 17.07$, $p < .001$). **Participants in the "partial" and "full CV support" groups reported lower physical demand than participants in the "no CV support" group** ("no CV support" group: $Mdn = 71.53$, IQR = 15.52; "partial CV support": $Mdn = 51.30$, IQR = 11.96; "full CV support": $Mdn = 38.60$, IQR = 13.20; Mann-Whitney $U = 21.00$, $p = .002$; $U = 10.00$, $p < .001$). **The "full CV support" group also reported significantly lower physical demand scores compared to the "partial CV support" group** (Mann-Whitney $U = 36.50$, $p = .04$). Furthermore, the subscale self-reported *performance* scores showed a significant effect across the three groups (Kruskal–Wallis $\chi^2(2) = 8.04$, $p = .02$). **The self-reported performance level was greater in the "partial" and "full CV support" groups compared to the group without CV support** ("no CV support" group: $Mdn = 38.64$, IQR = 14.36; "partial CV support" group: $Mdn = 44.72$, IQR = 15.71; "full CV support": $Mdn = 63.23$, IQR = 11.90; Mann-Whitney $U = 33.00$, $p = .02$; $U = 27.50$, $p = .01$). The performance subscale is directionally opposite to the workload subscales [34], with higher values indicating better perceived performance. Moreover, we found that *frustration* differed significantly among the groups (Kruskal–Wallis $\chi^2(2) = 11.98$, $p = .002$). **Designers reported significantly lower frustration in the "full CV support" group compared to both the "no CV support" and "partial CV support" groups** ("no CV support" group: $Mdn = 57.52$, IQR = 17.89; "partial CV support": $Mdn = 38.74$, IQR = 12.30; "full CV support": $Mdn = 32.71$, IQR = 15.62; Mann-Whitney $U = 21.500$, $p = .002$; $U = 22.00$, $p = .003$). There were no significant effects in the reported subscale *temporal demand* and *effort* across the three designer groups (Kruskal–Wallis $\chi^2(2) = 1.51$, $p = .47$; $\chi^2(2) = 0.71$, $p = .70$). **Overall, the results show that designers in the "full" and "partial CV support" groups reported lower TLX scores. This suggests that automation support in *MarkupLens* reduced their cognitive load in the given VBD task.**

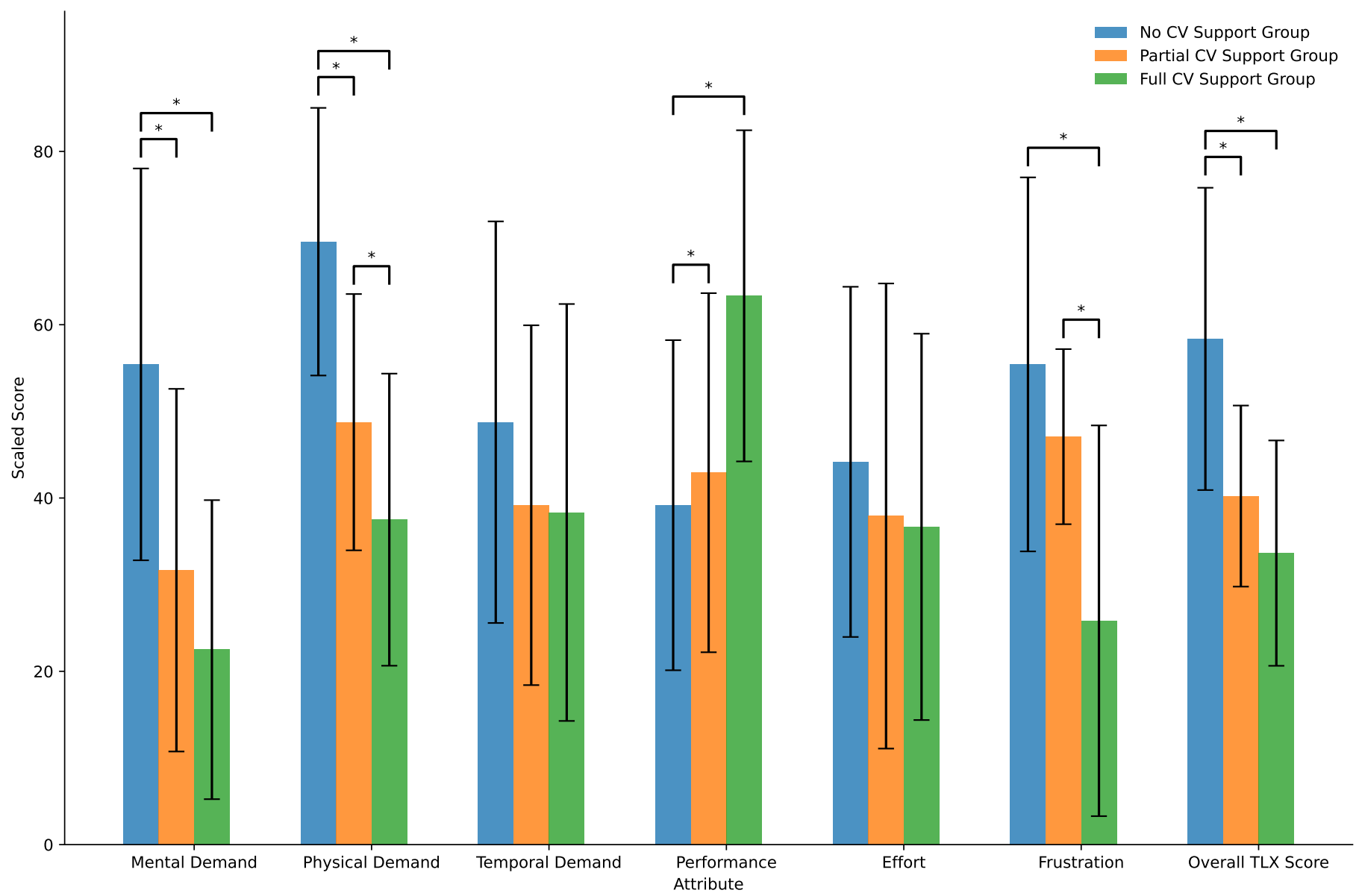

Fig. 4. Median values (*Mdn*) with interquartile range (IQR) error bars for the six NASA-TLX subscale and weighted overall scores across the three groups with different levels of automatic support. We observed main effects on the overall NASA-TLX score, as well as on the subscale scores for mental and physical demand, performance, and frustration. Note that higher scores on Performance indicate better perceived performance. Pairs with significant differences are marked with asterisks (*).

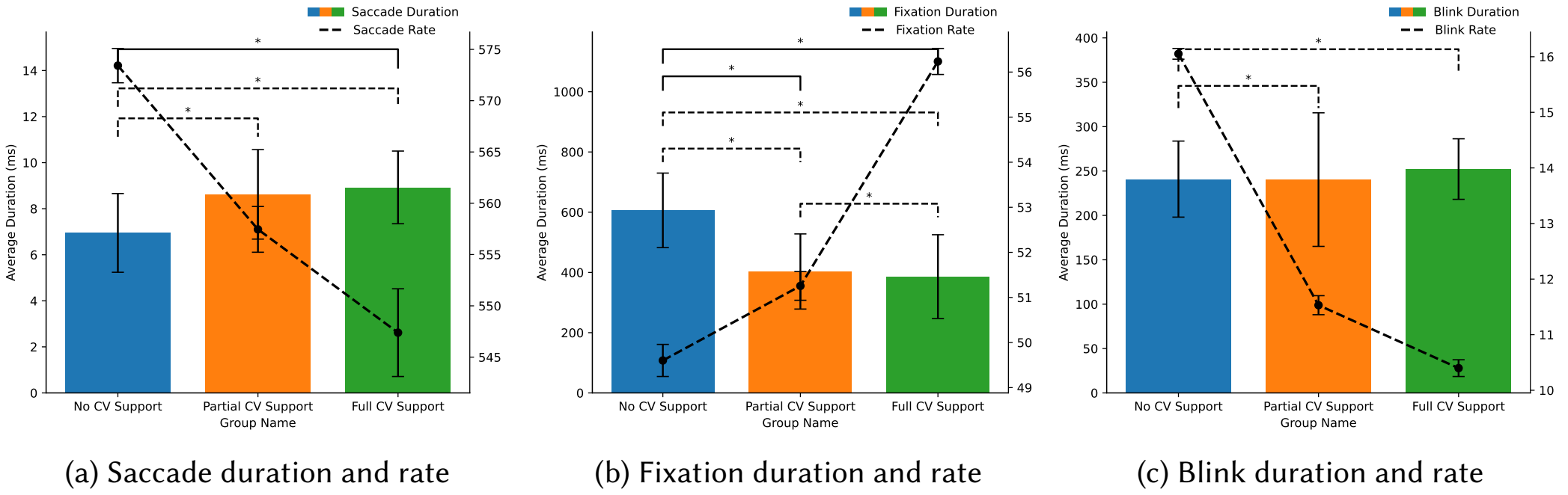

(a) Saccade duration and rate

(b) Fixation duration and rate

(c) Blink duration and rate

Fig. 5. Average values of eye-tracking metrics within the AOI, including saccadic movements (5a), fixations (5b), and blinks (5c), accompanied by SDs presented in error bars. Left y-axes in three plots denote scales for the average duration of events, while the right y-axes denote the event frequencies. We observed significant main effects in several metrics across the three groups: saccade duration, saccade rate, fixation duration, fixation rate, and blink rate. Pairs that demonstrated significant differences in duration are marked with solid lines with asterisks (*), while pairs with significant differences in rate are indicated by dashed lines with asterisks.

*5.3.2 Eye gaze.* Fig. 5 presents the saccade movements, fixation activities, and blink patterns of participants within the AOI across the three groups during the task. We observed significant differences in both the average duration and rate of saccades during tasks across the three groups (ANOVA $F_{2,33} = 4.07$, $p = .03$, $\eta^2 = .198$; $F_{2,33} = 218.28$, $p < .001$, $\eta^2 = .93$). On average, **designers in the "full CV support" group exhibited significantly longer saccade duration compared to those without CV support** (Bonferroni $p = .04$). While there was no significant difference in saccade duration between the "no" and "partial CV support" groups (Bonferroni $p = .10$), and between the "partial" and "full CV support" groups (Bonferroni $p = 1.00$), **we observed significantly lower average saccade rate in both the "full" and "partial CV support" groups compared to "no CV support" group** (Bonferroni $p < .001$; $p < .001$). Furthermore, we noted a significant variation in average fixation duration and rate across the three groups (ANOVA $F_{2,33} = 9.15$, $p < .001$, $\eta^2 = .36$; $F_{2,33} = 1259.36$, $p < .001$, $\eta^2 = .98$). Subsequent pairwise tests revealed that, compared to the "no CV support" group, designers in the "partial" and "full CV support" groups had significantly shorter average fixation duration (Bonferroni $p < .01$; $p < .01$). Additionally, the average fixation rates were significantly higher between the "no" and "partial", "partial" and "full", as well as between the "no" and "full CV support" groups (Bonferroni $p < .001$, $p < .001$; $p < .001$). In addition to fixation movements, we also found a significant difference in the average blink rates across the three groups (ANOVA $F_{2,33} = 4811.71$, $p < .001$, $\eta^2 = .99$), even though average blink duration remained relatively stable (ANOVA $F_{2,33} = 0.17$, $p = .84$, $\eta^2 = .01$). Pairwise analysis showed **designers in the "partial" and "full CV support" groups blinked on average less frequently during the task compared to the "no CV support" group** (Bonferroni $p < .001$; $p < .001$). **Overall, eye-movement results also indicated that designers in the "full" and "partial CV support" groups experienced lower cognitive load.**

### 5.4 Technology Acceptance and UX Analysis (RQ3)

*5.4.1 TAM.* We used the TAM questionnaire scores as a proxy for measuring participants' technology acceptance [17, 65] toward different levels of automated annotation in the VBD task. In Fig. 6, we gathered data on twelve attributes (see details in Table 3) and averaged the values into two categories: Perceived Usefulness (PU) and Perceived Ease-of-Use (PEU). From PU, we found a significant effect among the three groups (Kruskal-Wallis $\chi^2(2) = 6.70$, $p = .04$). Specifically,

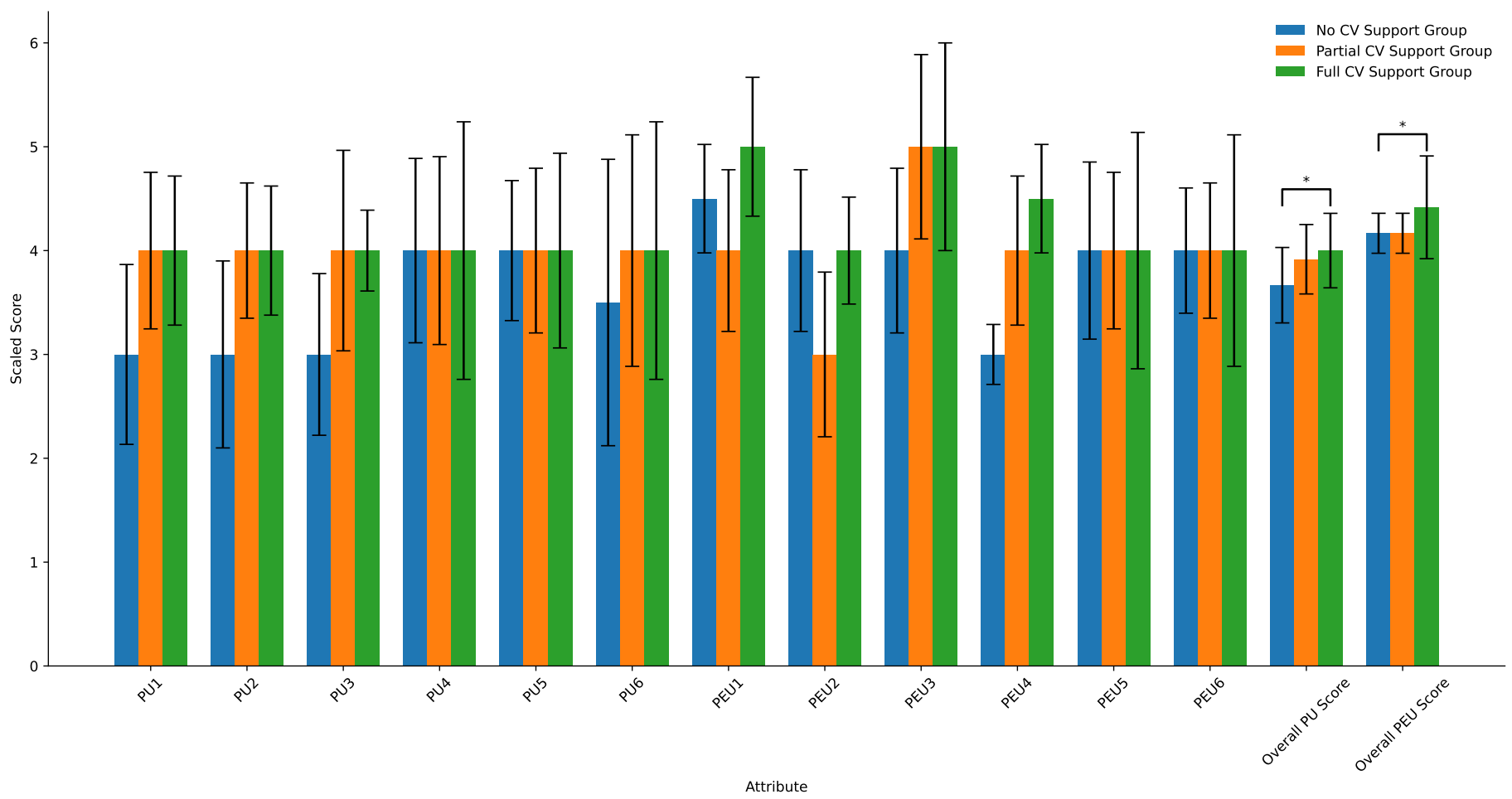

Fig. 6. Medians (*Mdn*) of TAM scores in three groups with interquartile range (IQR) indicated by error bars. Refer to Table 3 for the complete list of TAM statements matched with their corresponding attribute codes. We found that the overall Perceived Usefulness (PU) and Perceived Ease of Use (PUE) scores were significantly higher in the "full CV support" group compared to the "no CV support" group. All significant differences are marked with asterisks (*).

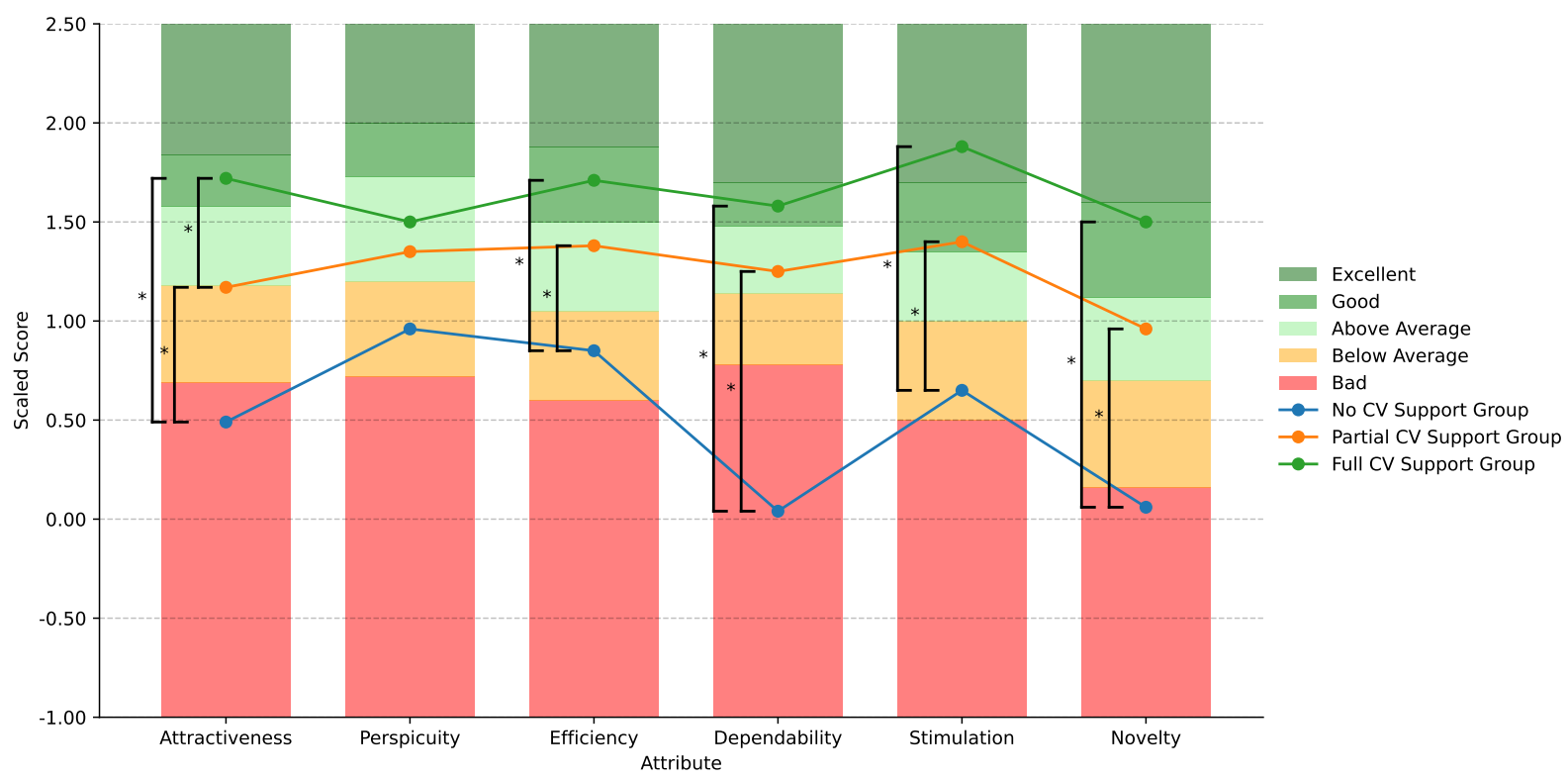

Fig. 7. The scaled average values for the six UEQ attributes are presented: Attractiveness, Perspicuity, Efficiency, Dependability, Stimulation, and Novelty. The segments are color-coded to represent performance benchmark values: less than 25% (bad), 25–50% (below average), 50–75% (above average), 75–90% (good), and greater than 90% (excellent) [58, 59]. We noted significant main effects in attributes including Attractiveness, Efficiency, Dependability, Stimulation, and Novelty. Pairs with significant differences are marked with asterisks (*).

**the median PU score was higher in the "full CV support" group compared to the "no CV support" group** ("full CV support" group: $Mdn = 4.13$, IQR = 0.84; "no CV support" group: $Mdn = 3.51$, IQR = 0.90; Mann-Whitney $U = 30.00$, $p = .02$). Additionally, we observed a main effect from the overall PEU scores of the three groups (Kruskal-Wallis $\chi^2(2) = 6.64$, $p = .04$). In pairwise evaluations, **the PEU score of the "full CV support" group were significantly higher than the "no CV support" group** ("full CV support" group: $Mdn = 4.35$, IQR = 0.70; "no CV support" group: $Mdn = 4.02$, IQR = 0.21; Mann-Whitney $U = 33.00$, $p = .03$).

*5.4.2 UEQ.* We applied a benchmark proposed by Schrepp et al. [58, 59] to transform 26 adjective pairs into 6 UX categories: Attractiveness, Perspicuity, Efficiency, Dependability, Stimulation, and Novelty. As shown in Fig. 7, the average attractiveness scores differed significantly across the three groups (ANOVA $F_{2,33} = 18.74$, $p < .001$, $\eta^2 = .53$). Specifically, the "partial CV support" group exhibited a significantly higher average score of attractiveness ($Mn = 1.17$, $SD = 0.40$) compared to the "no CV support" group ($Mn = 0.49$, $SD = 0.65$; Bonferroni $p < .01$). Participants in the "full CV support" group also reported higher **attractiveness** scores ($Mn = 1.72$, $SD = 0.40$) compared to both in the "no" CV" and "partial CV support" groups (Bonferroni $p < .001$; $p = .03$). No significant main effect was found in the average perspicuity scores (ANOVA $F_{2,33} = 2.25$, $p = .12$, $\eta^2 = .30$). However, further analysis revealed a significant disparity in efficiency scores among the three groups (ANOVA $F_{2,33} = 9.82$, $p < .001$, $\eta^2 = .37$). In pairwise tests, the groups with "full CV" ($Mn = 1.71$, $SD = 0.32$) and "partial CV support" ($Mn = 1.38$, $SD = 0.60$) outperformed the "no CV support" group ($Mn = 0.85$, $SD = 0.47$) in terms of average **efficiency** scores (Bonferroni $p < .001$; $p = .04$). For dependability, a significant effect was observed among the average scores (ANOVA $F_{2,33} = 24.48$, $p < .001$, $\eta^2 = .60$). Specifically, the average **dependability** scores for both the "partial" ($Mn = 1.25$, $SD = 0.61$) and "full CV" ($Mn = 1.58$, $SD = 0.53$) groups were significantly higher than "no CV support" group ($Mn = 0.04$, $SD = 0.56$, Bonferroni $p < .001$; $p < .001$). A significant effect was also observed in the average scores of stimulation (ANOVA $F_{2,33} = 13.91$, $p < .001$, $\eta^2 = .46$). Participants in both "full CV" ($Mn = 1.88$, $SD = 0.47$) and "partial CV support" ($Mn = 1.40$, $SD = 0.46$) groups reported higher average **stimulation** scores compared to "no CV support" group ($Mn = 0.65$, $SD = 0.75$; Bonferroni $p < .001$; $p < .01$). Lastly, our analysis revealed a significant main effect in the average novelty scores across the three groups (ANOVA $F_{2,33} = 19.43$, $p < .001$, $\eta^2 = .54$). Similarly, participants in both the "full CV" ($Mn = 1.50$, $SD = 0.54$) and "partial CV support" ($Mn = 0.96$, $SD = 0.40$) groups exhibited significantly higher average **novelty** scores compared to "no CV support" group ($Mn = 0.06$, $SD = 0.72$; Bonferroni $p < .001$; $p < .01$). **These results indicate that designers in the "partial CV" and "full CV support" experienced higher UX in terms of attractiveness, efficiency, dependability, stimulation and novelty than in "no CV support" group.**

## 6 Discussion

Our results suggest that more VBD-specialized automatic annotation support through *MarkupLens* improves manual annotation quality, reduces cognitive workload, and enhances multiple aspects of user experience. These findings indicate that practitioners could benefit from *MarkupLens* to improve video annotations and streamline the VBD workflow.

### 6.1 Higher level of automatic annotation support enhances critical thinking in VBD video annotation

Numerous prior studies have corroborated the role of CV-enhanced automation in improving efficiency and decision-making across diverse disciplines [53, 68]. In light of this, we were curious to explore whether the incorporation of automatic annotation support in *MarkupLens* could help designers to generate a higher quality of annotations in VBD tasks (RQ1). Our findings suggest that designers tended to produce fewer VBD annotations but more detailed textual content in their texts in "partial CV" and "full CV support" groups compared to "no CV support" group. With higher levels of automatic support, designers demonstrated more analytical reasoning and deeper insights into the VBD cases. This capability is essential for critical thinking in the video exploration and ideation process as it increases the possibility of generating comprehensive and reflective insights. Although practitioners in the "no CV support" group produced more annotations than those in the "partial CV" and "full CV support" groups, these annotations were predominantly limited to objective names or brief descriptions of observed elements. A similar pattern was also seen in the "partial

CV support" group. Moreover, practitioners in the "partial CV" and "full CV support" groups produced more frequent in-depth textual annotations and critical insights. These findings suggest that specialized automatic annotation in VBD videos supports a more comprehensive grasp of design elements and their enriched interactions with the given context. **The enhanced automation in VBD video annotations improves practitioners' learning experience and helps having deeper interaction with the content.** Notably, higher-quality design annotations suggest that practitioners are developing stronger problem-solving skills and creative thinking. These findings demonstrate the practical impact of specialized design tasks across contexts and point to the need for future research on exploring appropriate reliance on automation in diverse tasks [42, 48].

### 6.2 Higher level of automatic annotation support reduces cognitive load in VBD task

The NASA-TLX results show that practitioners with automatic VBD annotation support ("partial CV" and "full CV support" groups) had lower cognitive load in the VBD task compared to the human-only group ("no CV support"). Specifically, participants in the "partial CV" and "full CV support" groups had lower mental demand, physical demand, and frustration. At the same time, they experienced higher task performance in conditions with support in VBD annotations. Together, these results suggest [30] that higher levels of automatic annotation support can reduce practitioners' cognitive load in VBD tasks.

That said, we still wanted to investigate practitioners' eye-movement activity as a window into their thinking processes [57]. In conditions with a high level of automatic annotation support (the "partial CV" and "full CV support" groups), practitioners exhibited longer but less frequent saccadic movements, which also serves as an objective indicator of reduced cognitive load. [6, 27]. In examining fixations, our analysis revealed that practitioners with higher automatic support exhibited shorter fixation durations when engaging with VBD videos compared to other conditions. Shorter fixations are associated with decreased cognitive load [55] and suggest that practitioners were able to more efficiently process visual information [16]. Additionally, a higher frequency of fixations in eye movements is commonly associated with reduced cognitive load [14, 19]. We observed this pattern in conditions with more automatic support, which suggests that practitioners spent less effort on processing visual content in the VBD videos and could allocate more cognitive resources to higher-level thinking. In turn, the reduced cognitive load enables practitioners to focus more intently on the VBD task at hand and retain greater capacity to creative and critical thinking [56]. Additionally, in the groups with higher levels of automatic support, we observed a reduced blinking rate in eye movements. This pattern is associated with lower cognitive load [8, 23, 47]. In practice, decreased blink frequency is associated with reduced mental fatigue, fewer attention lapses, and lower stress [12, 64]. Reduced mental fatigue, in turn, enhances practitioners' overall task productivity [9, 56]. As such, practitioners with higher automatic support in VBD task can have a more relaxed mental state with less stress and fatigue, which contributed to a reduction in cognitive load (RQ2).

In this study, we used an urban biking traffic case from urban design to illustrate how *MarkupLens* can support the VBD workflow. This scenario was selected not only for its accessibility and relevance to the local community but also because it provided a sufficient number of events, interruptions, and objects within a short study period. For example, practitioners could identify patterns in cyclist behavior, analyze interactions between cyclists and vehicles, and evaluate the impact of roadway conditions and situational elements on traffic flow. We believe our findings here highlight how *MarkupLens* contributes to the VBD workflow by reducing the cognitive workload associated with repetitive tasks, such as searching for basic behavior information and simple objects of interest. This reduction allows practitioners to focus on interpreting and analyzing higher-order design insights.

### 6.3 Higher level of automatic annotation support enhances UX of VBD practitioners

To better understand VBD practitioners' perceptions of automation support, we examined answers from TAM and UEQ questionnaires. TAM scores provide understandings into technology acceptance and perceived ease of use [17], while UEQ evaluates users' interactions with *MarkupLens* across six attributes [58, 59]. Overall, designers in groups with higher levels of automatic support reported greater Perceived Usefulness (PU) and Perceived Ease of Use (PEU) toward *MarkupLens*. From the think-aloud recordings during the experiment, practitioners in groups with higher levels of automatic support also noted that the consistency of automatically generated annotations, such as color-coded instance segmentation, made it easier and quicker to find the content they wanted to search for. We believe this explains why higher UX scores were observed under higher automatic support conditions, as automatic annotation improved visual coherence and usability in the VBD task. Participants in higher automatic support groups highlighted these features as particularly helpful for maintaining focus and understanding complex visual scenes.

Furthermore, we observed higher levels of attractiveness, efficiency, dependability, stimulation, and novelty in conditions with greater automatic annotation support (UEQ). We believe that practitioners regarded the automatic support in the VBD annotation as a dependable resource for case analysis (dependability), which facilitated the process and encouraged the generation of more related insights (efficiency and stimulation). Additionally, participants may have perceived the provided automatic annotation support as adding interest to the manual VBD annotation process (novelty). Taken together, results from both TAM and UEQ suggest that designers experienced a better user experience in conditions with greater automatic support in VBD tasks (RQ3). **The higher UX ensures intuitive navigation and interaction for designer when handling videos. It reduces the time and effort required to learn and operate our *MarkupLens* for VBD tasks.** This enhanced UX can also allow practitioners to focus more on self-expressed creativity rather than grappling with the layout of the interface in design exploration and ideation [26].

### 6.4 Implications from *MarkupLens* for Human–AI Collaboration

We conceptualize CV support as a *continuum of automation* including manual annotation, partially and fully automated assistance. Treating support levels as points on this spectrum clarifies trade-offs between efficiency and reflective engagement. In practice, higher assistance in the study can free practitioners' attention for higher-level interpretation, yet excessive automation risks narrowing designers' exploration or inducing over-reliance.

Our findings suggest three design principles for adjustable autonomy in using AI-supported tools for VBD tasks. **P1—Expose controllability.** Provide explicit, low-friction controls for model confidence and assistance level so designers can calibrate support to the task moment. **P2—Maintain agency with reversible operations.** Automated suggestions should be easy to accept, edit, or dismiss, preserving human authorship in video exploration. **P3—Make provenance legible.** Visually differentiate human vs. machine-generated annotations and reveal rationales (e.g., category, confidence), helping practitioners gain trust without cognitive overhead.

### 6.5 Limitations

In this study, we used a VBD task from urban design to assess whether and how specialized automatic annotation support through *MarkupLens* can assist practitioners' manual annotation in VBD. We conducted a controlled between-subjects lab study with three levels of automatic support. However, several limitations must be acknowledged. Design scenarios using videos can vary greatly, from camera angles to content such as interviews, outdoor operations, and indoor activities [70]. This diversity can affect how the annotation process is quantified. In our study, we evaluated *MarkupLens* with a single scenario: analyzing a video of a cyclist biking through traffic to design safer commuting policies. This limitation may introduce bias by not accounting for other design cases. Nevertheless, the selected task represents a realistic case of VBD in urban planning and traffic engineering [39, 60]. Moreover, designers' interpretations are never totally under their control because they frequently rely on personal past experiences during the VBD process [70]. The inherent variability also suggests that no single scenario can cover all possible cases, even though this shows that our chosen case is sufficiently representative for evaluating automatic annotation in actual VBD practices.

Additionally, the aims of different VBD tasks are inherently varied [70]. In real-world settings, designers may pursue a range of objectives when engaging with VBD videos: from understanding customers' needs to identifying practical challenges or seeking solutions [69]. Each of these aims introduces further complexity to the study. However, we created a workflow and provided many options for customizing automatic support in *MarkupLens* as illustrated in Fig. 2a. This provides future practitioners with the flexibility to select options based on their specific VBD tasks. In addition, to control the variables in our study while balancing computational resources, we limited our testing to one model (YOLOv8x) and used instance segmentation for automatic annotation visualization. Nevertheless, at the time of writing, AI companies such as Ultralytics[14] and Meta[15] had already introduced more advanced models for automatic detection in videos (e.g., YOLO11[16] and SAM 2[17]). We acknowledge that our adopted detection models may now be considered outdated, but they still demonstrate that providing a level of automation support can assist VBD practitioners' thinking and reduce their effort when processing information from design videos. We also believe this topic remains valuable, as future research should examine whether more powerful models further impact the VBD workflow in line with our findings.

## 7 Conclusion

The rapid progress of automatic detection from CV provides the potential to enhance and scale up the video-based design (VBD) practice. Earlier studies have applied automatic detection models to professional annotation tasks in other specialized domains such as medical imaging. In this paper, we explored the use of automatic detection models to support VBD practitioners in video analysis. We developed *MarkupLens*, an CV-powered prototype that provides automatic annotation support to designers, and validated it through a VBD task from urban design. The design task was evaluated across three conditions with varying levels of automatic annotation. We investigated how CV-supported automatic annotation support changes manual annotation practices and workload in design practices, which are foundational conditions that enable reflection and insight generation in VBD tasks. Our results show that automatic annotation support not only improves practitioners' manual annotation quality but also reduces cognitive load and enhances user experience with *MarkupLens*. We argue that refined automatic detection has substantial potential to scale up video analysis in VBD. Additionally, by framing assistance levels as a spectrum of human–AI collaboration, we offer actionable guidance for adjustable autonomy and mixed-initiative design in VBD. We believe the approach might extend to general strategies for scaling up analysis for other data-sensitive contexts where people and intelligent systems jointly interpret complex video.

[14]https://www.ultralytics.com/ (last accessed: November 19, 2025).
[15]https://ai.meta.com/ (last accessed: November 19, 2025).
[16]https://docs.ultralytics.com/models/yolo11/ (last accessed: November 19, 2025).
[17]https://ai.meta.com/sam2/ (last accessed: November 19, 2025).

## References

[1] Nantheera Anantrasirichai and David Bull. 2022. Artificial intelligence in the creative industries: a review. *Artificial intelligence review* 55, 1 (2022), 589–656.

[2] Onur Asan and Enid Montague. 2014. Using video-based observation research methods in primary care health encounters to evaluate complex interactions. 21, 4 (2014), 161–170. doi:10.14236/jhi.v21i4.72

[3] Hasan Ayaz, Patricia A. Shewokis, Scott Bunce, Kurtulus Izzetoglu, Ben Willems, and Banu Onaral. 2012. Optical brain monitoring for operator training and mental workload assessment. 59, 1 (2012), 36–47. doi:10.1016/j.neuroimage.2011.06.023

[4] Fabio Babiloni. 2019. Mental Workload Monitoring: New Perspectives from Neuroscience. In *Human Mental Workload: Models and Applications* (Cham) *(Communications in Computer and Information Science)*, Luca Longo and Maria Chiara Leva (Eds.). Springer International Publishing, 3–19. doi:10.1007/978-3-030-32423-0_1

[5] Laura Baecher and Bede McCormack. 2015. The impact of video review on supervisory conferencing. 29, 2 (2015), 153–173. doi:10.1080/09500782.2014.992905

[6] Mahnaz Behroozi, Alison Lui, Ian Moore, Denae Ford, and Chris Parnin. 2018. Dazed: measuring the cognitive load of solving technical interview problems at the whiteboard. In *Proceedings of the 40th International Conference on Software Engineering: New Ideas and Emerging Results* (Gothenburg Sweden). ACM, 93–96. doi:10.1145/3183399.3183415

[7] Simone Bianco, Gianluigi Ciocca, Paolo Napoletano, and Raimondo Schettini. 2015. An interactive tool for manual, semi-automatic and automatic video annotation. 131 (2015), 88–99. doi:10.1016/j.cviu.2014.06.015

[8] Francesco N. Biondi, Babak Saberi, Frida Graf, Joel Cort, Prarthana Pillai, and Balakumar Balasingam. 2023. Distracted worker: Using pupil size and blink rate to detect cognitive load during manufacturing tasks. 106 (2023), 103867. doi:10.1016/j.apergo.2022.103867

[9] Maarten A. S. Boksem and Mattie Tops. 2008. Mental fatigue: Costs and benefits. 59, 1 (2008), 125–139. doi:10.1016/j.brainresrev.2008.07.001

[10] Abhigna B.s., Nitasha Soni, and Shilpa Dixit. 2018. Crowdsourcing – A Step Towards Advanced Machine Learning. 132 (2018), 632–642. doi:10.1016/j.procs.2018.05.062

[11] Zana Buçinca, Maja Barbara Malaya, and Krzysztof Z. Gajos. 2021. To Trust or to Think: Cognitive Forcing Functions Can Reduce Overreliance on AI in AI-assisted Decision-making. 5 (2021), 188:1–188:21. Issue CSCW1. doi:10.1145/3449287

[12] Philipp P. Caffier, Udo Erdmann, and Peter Ullsperger. 2003. Experimental evaluation of eye-blink parameters as a drowsiness measure. 89, 3 (2003), 319–325. doi:10.1007/s00421-003-0807-5

[13] Zhe Cao, Gines Hidalgo, Tomas Simon, Shih-En Wei, and Yaser Sheikh. 2021. OpenPose: Realtime Multi-Person 2D Pose Estimation Using Part Affinity Fields. *IEEE Transactions on Pattern Analysis and Machine Intelligence* 43, 1 (2021), 172–186. doi:10.1109/TPAMI.2019.2929257

[14] Siyuan Chen, Julien Epps, Natalie Ruiz, and Fang Chen. 2011. Eye activity as a measure of human mental effort in HCI. In *Proceedings of the 16th international conference on Intelligent user interfaces* (New York, NY, USA) *(IUI '11)*. Association for Computing Machinery, 315–318. doi:10.1145/1943403.1943454

[15] Valerie Chen, Q. Vera Liao, Jennifer Wortman Vaughan, and Gagan Bansal. 2023. Understanding the Role of Human Intuition on Reliance in Human-AI Decision-Making with Explanations. 7 (2023), 370:1–370:32. Issue CSCW2. doi:10.1145/3610219

[16] Charles Clifton, Fernanda Ferreira, John M. Henderson, Albrecht W. Inhoff, Simon P. Liversedge, Erik D. Reichle, and Elizabeth R. Schotter. 2016. Eye movements in reading and information processing: Keith Rayner's 40year legacy. 86 (2016), 1–19. doi:10.1016/j.jml.2015.07.004

[17] Fred D. Davis. 1989. Perceived Usefulness, Perceived Ease of Use, and User Acceptance of Information Technology. 13, 3 (1989), 319–340. doi:10.2307/249008 Publisher: Management Information Systems Research Center, University of Minnesota.

[18] Fred D. Davis, Richard P. Bagozzi, and Paul R. Warshaw. 1989. User Acceptance of Computer Technology: A Comparison of Two Theoretical Models. 35, 8 (1989), 982–1003. doi:10.1287/mnsc.35.8.982 Publisher: INFORMS.

[19] M. De Rivecourt, M. N. Kuperus, W. J. Post, and L. J.M. Mulder. 2008. Cardiovascular and eye activity measures as indices for momentary changes in mental effort during simulated flight. 51, 9 (2008), 1295–1319. doi:10.1080/00140130802120267

[20] Dazhen Deng, Jiang Wu, Jiachen Wang, Yihong Wu, Xiao Xie, Zheng Zhou, Hui Zhang, Xiaolong Zhang, and Yingcai Wu. 2021. Eventanchor: Reducing human interactions in event annotation of racket sports videos. In *Proceedings of the 2021 CHI Conference on Human Factors in Computing Systems*. 1–13.

[21] Yuanyuan Ding, Junchi Yan, Guoqiang Hu, and Jun Zhu. 2022. Cognitive visual inspection service for LCD manufacturing industry. In *2022 5th International Conference on Pattern Recognition and Artificial Intelligence (PRAI)*. IEEE, 329–336.

[22] Daniel E. Ehrmann, Sara N. Gallant, Sujay Nagaraj, Sebastian D. Goodfellow, Danny Eytan, Anna Goldenberg, and Mjaye L. Mazwi. 2022. Evaluating and reducing cognitive load should be a priority for machine learning in healthcare. 28, 7 (2022), 1331–1333. doi:10.1038/s41591-022-01833-z Number: 7 Publisher: Nature Publishing Group.

[23] Vérane Faure, Régis Lobjois, and Nicolas Benguigui. 2016. The effects of driving environment complexity and dual tasking on drivers' mental workload and eye blink behavior. 40 (2016), 78–90. doi:10.1016/j.trf.2016.04.007

[24] Ana Fucs, Juliana Jansen Ferreira, Vinícius C. V. B. Segura, Beatriz de Paulo, Rogerio Abreu De Paula, and Renato Cerqueira. 2020. Sketch-based Video A Storytelling for UX Validation in AI Design for Applied Research. *CHI Extended Abstracts* (2020). doi:10.1145/3334480.3375221

[25] Eshan Gaur, Vikas Saxena, and Sandeep K Singh. 2018. Video annotation tools: A Review. In *2018 International Conference on Advances in Computing, Communication Control and Networking (ICACCCN)*. 911–914. doi:10.1109/ICACCCN.2018.8748669

[26] Michail Giannakos and Ioannis Leftheriotis. 2015. How Space and Tool Availability Affect User Experience and Creativity in Interactive Surfaces?. In *Proceedings of the 2015 ACM SIGCHI Conference on Creativity and Cognition* (New York, NY, USA) *(C&C '15)*. Association for Computing Machinery, 201–204. doi:10.1145/2757226.2764554

[27] Joseph H Goldberg and Xerxes P Kotval. 1999. Computer interface evaluation using eye movements: methods and constructs. 24, 6 (1999), 631–645. doi:10.1016/S0169-8141(98)00068-7

[28] Canadian Centre for Occupational Health Government of Canada and Safety. 2023. *CCOHS: Office Ergonomics - Positioning the Monitor.* https://www.ccohs.ca/oshanswers/ergonomics/office/monitor_positioning.html Last Modified: 2023-06-13.

[29] Carl Gutwin, Scott Bateman, Gaurav Arora, and Ashley Coveney. 2017. Looking Away and Catching Up: Dealing with Brief Attentional Disconnection in Synchronous Groupware. In *Proceedings of the 2017 ACM Conference on Computer Supported Cooperative Work and Social Computing* (Portland Oregon USA). ACM, 2221–2235. doi:10.1145/2998181.2998226

[30] Sandra G. Hart and Lowell E. Staveland. 1988. Development of NASA-TLX (Task Load Index): Results of Empirical and Theoretical Research. In *Advances in Psychology*. Vol. 52. Elsevier, 139–183. doi:10.1016/S0166-4115(08)62386-9

[31] Tianhao He, Andrija Stanković, Evangelos Niforatos, and Gerd Kortuem. 2025. DesignMinds: Enhancing Video-Based Design Ideation with a Vision-Language Model and a Context-Injected Large Language Model. In *Proceedings of the 7th ACM Conference on Conversational User Interfaces*. 1–15.

[32] John M. Hinson, Tina L. Jameson, and Paul Whitney. 2003. Impulsive decision making and working memory. 29, 2 (2003), 298–306. doi:10.1037/0278-7393.29.2.298

[33] Kenneth Holmqvist, Marcus Nyström, Richard Andersson, Richard Dewhurst, Halszka Jarodzka, and Joost Van de Weijer. 2011. *Eye tracking: A comprehensive guide to methods and measures.* oup Oxford.

[34] Human Performance Research Group at NASA's Ames Research Center. 2022. *NASA Task Load Index (NASA-TLX) Paper and Pencil Version Instruction Manual.* NASA, Moffett Field, CA. Available at: https://humansystems.arc.nasa.gov/groups/tlx/tlxpaperpencil.php.

[35] Robert Jenssen, Davide Roverso, et al. 2019. Intelligent monitoring and inspection of power line components powered by UAVs and deep learning. *IEEE Power and energy technology systems journal* 6, 1 (2019), 11–21.

[36] Vaiva Kalnikaitė, Patrick Ehlen, and Steve Whittaker. 2012. Markup as you talk: establishing effective memory cues while still contributing to a meeting. In *Proceedings of the ACM 2012 conference on Computer Supported Cooperative Work* (Seattle Washington USA). ACM, 349–358. doi:10.1145/2145204.2145260

[37] Isaak Kavasidis, Simone Palazzo, Roberto Di Salvo, Daniela Giordano, and Concetto Spampinato. 2014. An innovative web-based collaborative platform for video annotation. 70, 1 (2014), 413–432. doi:10.1007/s11042-013-1419-7

[38] Jeongyeon Kim, Daeun Choi, Nicole Lee, Matt Beane, and Juho Kim. 2023. Surch: Enabling structural search and comparison for surgical videos. In *Proceedings of the 2023 CHI Conference on Human Factors in Computing Systems*. 1–17.

[39] Alan Latham and Peter R H Wood. 2015. Inhabiting Infrastructure: Exploring the Interactional Spaces of Urban Cycling. 47, 2 (2015), 300–319. doi:10.1068/a140049p

[40] Tianbiao Liang, Tianyuan Liu, Junliang Wang, Jie Zhang, and Pai Zheng. 2024. Causal deep learning for explainable vision-based quality inspection under visual interference. *Journal of Intelligent Manufacturing* (2024), 1–22.

[41] Julie S Linsey, Ian Tseng, Katherine Fu, Jonathan Cagan, Kristin L Wood, and Christian Schunn. 2010. A study of design fixation, its mitigation and perception in engineering design faculty. (2010).

[42] Keming Liu. 2006. Annotation As an Index to Critical Writing. 41, 2 (2006), 192–207. doi:10.1177/0042085905282261 Publisher: SAGE Publications Inc.

[43] Yikun Liu, Yuan Jia, Wei Pan, and Mark S. Pfaff. 2014. Supporting task resumption using visual feedback. In *Proceedings of the 17th ACM conference on Computer supported cooperative work & social computing* (Baltimore Maryland USA). ACM, 767–777. doi:10.1145/2531602.2531710

[44] Wendy E Mackay. 2002. Using video to support interaction design. *DVD Tutorial, CHI* 2, 5 (2002), 1–42.

[45] Wendy E. Mackay, Anne V. Ratzer, and Paul Janecek. 2000. Video artifacts for design: bridging the Gap between abstraction and detail. In *Proceedings of the 3rd conference on Designing interactive systems: processes, practices, methods, and techniques* (New York City New York USA). ACM, 72–82. doi:10.1145/347642.347666

[46] Antonio Maffei and Alessandro Angrilli. 2019. Spontaneous blink rate as an index of attention and emotion during film clips viewing. 204 (2019), 256–263. doi:10.1016/j.physbeh.2019.02.037

[47] Alfonso Magliacano, Salvatore Fiorenza, Anna Estraneo, and Luigi Trojano. 2020-09-25. Eye blink rate increases as a function of cognitive load during an auditory oddball paradigm. 736 (2020-09-25), 135293. doi:10.1016/j.neulet.2020.135293

[48] Siddharth Mehrotra, Chadha Degachi, Oleksandra Vereschak, Catholijn M Jonker, and Myrthe L Tielman. 2023. A systematic review on fostering appropriate trust in human-AI interaction. *arXiv preprint arXiv:2311.06305* (2023).

[49] Xiangming Mu. 2010. Towards effective video annotation: An approach to automatically link notes with video content. 55, 4 (2010), 1752–1763. doi:10.1016/j.compedu.2010.07.021

[50] Tanmay Nath, Alexander Mathis, An Chi Chen, Amir Patel, Matthias Bethge, and Mackenzie Weygandt Mathis. 2019. Using DeepLabCut for 3D markerless pose estimation across species and behaviors. *Nature protocols* 14, 7 (2019), 2152–2176.

[51] Molly Jane Nicholas, Nicolai Marquardt, Michel Pahud, Nathalie Riche, Hugo Romat, Christopher Collins, David Ledo, Rohan Kadekodi, Badrish Chandramouli, and Ken Hinckley. 2023. Escapement: A tool for interactive prototyping with video via sensor-mediated abstraction of time. In

*Proceedings of the 2023 CHI Conference on Human Factors in Computing Systems*. 1–14.

[52] Marzan Tasnim Oyshi, Sebastian Vogt, and Stefan Gumhold. 2023. TmoTA: Simple, Highly Responsive Tool for Multiple Object Tracking Annotation. In *Proceedings of the 2023 CHI Conference on Human Factors in Computing Systems*. 1–11.

[53] Christoffer Bøgelund Rasmussen, Kristian Kirk, and Thomas B. Moeslund. 2022. The Challenge of Data Annotation in Deep Learning—A Case Study on Whole Plant Corn Silage. 22, 4 (2022), 1596. doi:10.3390/s22041596

[54] Keith Rayner. 1998. Eye movements in reading and information processing: 20 years of research. *Psychological bulletin* 124, 3 (1998), 372.

[55] Keith Rayner. 1998. Eye movements in reading and information processing: 20 years of research. 124, 3 (1998), 372–422. doi:10.1037/0033-2909.124.3.372 Place: US Publisher: American Psychological Association.

[56] Jenni L. Redifer, Christine L. Bae, and Qin Zhao. 2021. Self-efficacy and performance feedback: Impacts on cognitive load during creative thinking. 71 (2021), 101395. doi:10.1016/j.learninstruc.2020.101395

[57] Dario D Salvucci. 1999. Inferring intent in eye-based interfaces: tracing eye movements with process models. In *Proceedings of the SIGCHI conference on Human Factors in Computing Systems*. 254–261.

[58] Martin Schrepp, Andreas Hinderks, and Jörg Thomaschewski. 2014. Applying the User Experience Questionnaire (UEQ) in Different Evaluation Scenarios. 383–392. doi:10.1007/978-3-319-07668-3_37

[59] Martin Schrepp, Andreas Hinderks, and Jörg Thomaschewski. 2017. Construction of a Benchmark for the User Experience Questionnaire (UEQ). 4 (2017), 40–44. doi:10.9781/ijimai.2017.445

[60] Justin Spinney. 2011. A Chance to Catch a Breath: Using Mobile Video Ethnography in Cycling Research. 6, 2 (2011), 161–182. doi:10.1080/17450101.2011.552771

[61] John A Stern, Larry C Walrath, and Robert Goldstein. 1984. The endogenous eyeblink. *Psychophysiology* 21, 1 (1984), 22–33.

[62] Deborah Tatar. 1989. Using video-based observation to shape the design of a new technology. *ACM SIGCHI Bulletin* 21, 2 (1989), 108–111. doi:10.1145/70609.70628

[63] Russ Tedrake. 2023. *CH. 9 - Object Detection and Segmentation, Robotic Manipulation.* http://manipulation.mit.edu

[64] J. A. Veltman and A. W. K. Gaillard. 1998. Physiological workload reactions to increasing levels of task difficulty. 41, 5 (1998), 656–669. doi:10.1080/001401398186829 Publisher: Taylor & Francis eprint: https://doi.org/10.1080/001401398186829.

[65] Viswanath Venkatesh. 2000. Determinants of Perceived Ease of Use: Integrating Control, Intrinsic Motivation, and Emotion into the Technology Acceptance Model. 11, 4 (2000), 342–365. doi:10.1287/isre.11.4.342.11872 Publisher: INFORMS.

[66] Laurie Vertelney. 1989. Using video to prototype user interfaces. *ACM SIGCHI Bulletin* 21, 2 (1989), 57–61. doi:10.1145/70609.70615

[67] Athanasios Voulodimos, Nikolaos Doulamis, Anastasios Doulamis, and Eftychios Protopapadakis. 2018. Deep Learning for Computer Vision: A Brief Review. 2018 (2018), e7068349. doi:10.1155/2018/7068349

[68] Jan Witowski, Jongmun Choi, Soomin Jeon, Doyun Kim, Joowon Chung, John Conklin, Maria Gabriela Figueiro Longo, Marc D. Succi, and Synho Do. 2021. MarkIt: A Collaborative Artificial Intelligence Annotation Platform Leveraging Blockchain For Medical Imaging Research. 4 (2021), 10.30953/bhty.v4.176. doi:10.30953/bhty.v4.176

[69] Salu Ylirisku and Jacob Buur. 2007. Making sense and editing videos. In *Designing with video: Focusing the user-centred design process*. Springer London, 86—-135. doi:10.1007/978-1-84628-961-3_2

[70] Salu Ylirisku and Jacob Buur. 2007. Studying what people do. In *Designing with video: Focusing the user-centred design process*. Springer London, 36–85. doi:10.1007/978-1-84628-961-3_2

[71] Mingrui Ray Zhang, Mingyuan Zhong, and Jacob O Wobbrock. 2022. Ga11y: An automated gif annotation system for visually impaired users. In *Proceedings of the 2022 CHI Conference on Human Factors in Computing Systems*. 1–16.

[72] Tianyi Zhang, Abdallah El Ali, Chen Wang, Alan Hanjalic, and Pablo Cesar. 2020. Rcea: Real-time, continuous emotion annotation for collecting precise mobile video ground truth labels. In *Proceedings of the 2020 CHI Conference on Human Factors in Computing Systems*. 1–15.

[73] Zheng Zhang, Zheng Ning, Chenliang Xu, Yapeng Tian, and Toby Jia-Jun Li. 2023. PEANUT: A Human-AI Collaborative Tool for Annotating Audio-Visual Data. In *Proceedings of the 36th Annual ACM Symposium on User Interface Software and Technology* (New York, NY, USA) *(UIST '23)*. Association for Computing Machinery, 1–18. doi:10.1145/3586183.3606776

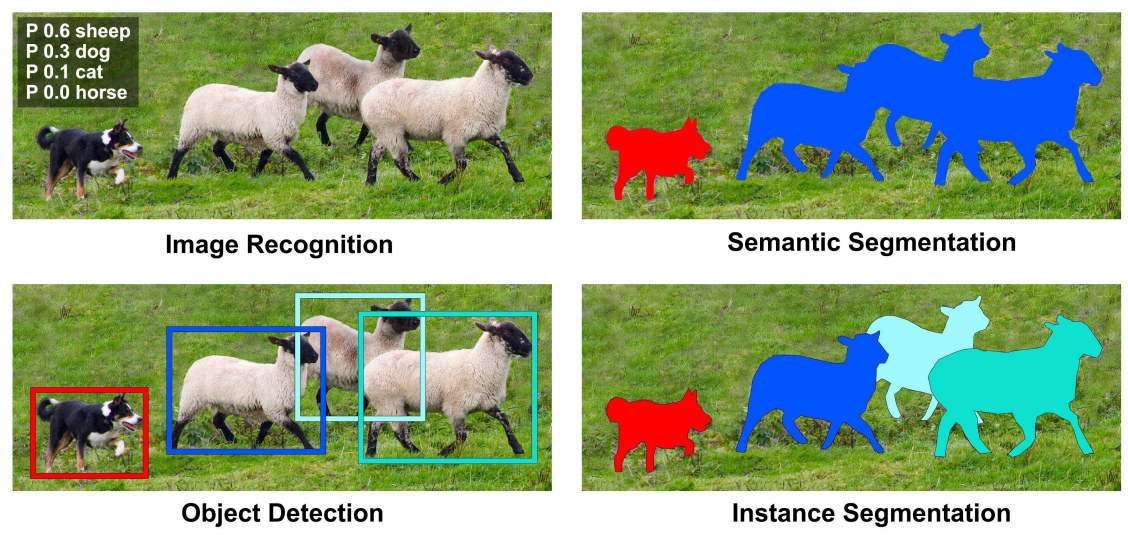

Fig. 8. An example demonstrating 4 commonly used annotation visualization [63]

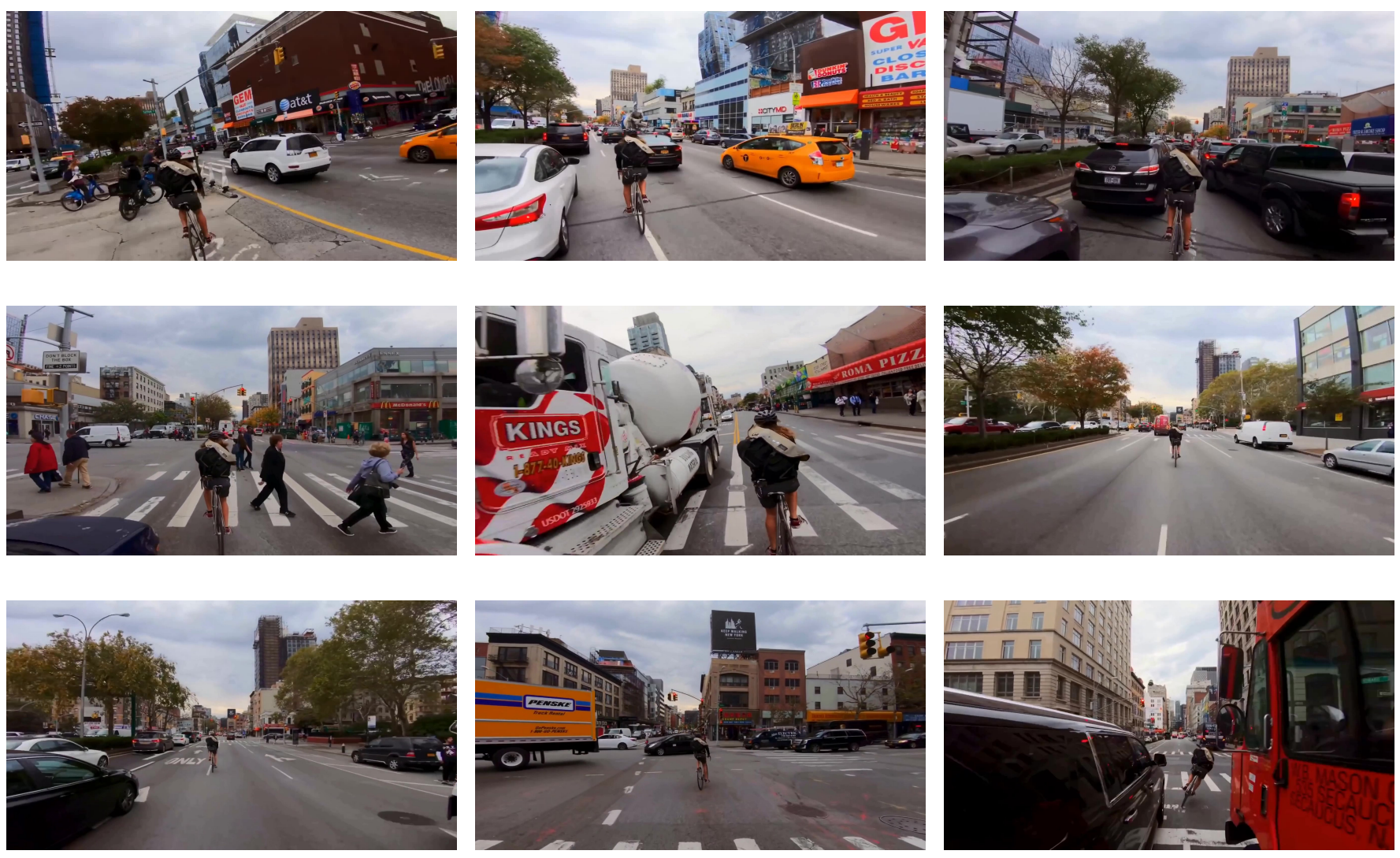

Fig. 9. Screenshots from the VBD video used in the study, where designers were asked to develop safer urban commuting policies for cyclists. Frames are shown at frame numbers 1, 601, 1201, 1801, 2401, 3001, 3601, 4201, and 4801.

Table 3. Statements of TAM in perspectives of Perceived Usefulness (PU) and Perceived Ease-of-Use (PEU). For simplicity, we coded each statement with a unique attribute code.

| Perspective | Statement | Attribute Code |
|---|---|---|
| Perceived Usefulness (PU) | Using the prototype helps me to accomplish the task more quickly. | PU1 |
| | Using the prototype improves my work performance. | PU2 |
| | Using the prototype improves my work productivity. | PU3 |
| | Using the prototype enhances my effectiveness at work. | PU4 |
| | Using the prototype makes it easier to do my work. | PU5 |
| | I find the prototype useful in my work. | PU6 |
| Perceived Ease-of-Use (PEU) | Learning to operate the annotation system has been easy for me. | PEU1 |
| | I find it easy to get the prototype to do what I want to do. | PEU2 |
| | My interaction with the prototype is clear and understandable. | PEU3 |
| | I find the prototype is flexible to interact with. | PEU4 |
| | It is easy for me to become skillful at using the prototype. | PEU5 |
| | I find the prototype easy to use. | PEU6 |